\documentclass[english,superscriptaddress,floatfix,twocolumn,oneside, amsmath,amssymb,amsfonts,aps,pra,floatfix]{revtex4-1}
\usepackage[english]{babel}
\usepackage[utf8x]{inputenc}
\usepackage{microtype}
\usepackage{graphicx}
\usepackage{siunitx}
\usepackage{xspace}
\usepackage[hidelinks]{hyperref}
\usepackage{bm}
\usepackage{bbold}
\usepackage{natbib}
\usepackage{csquotes}
\usepackage{qcircuit}

\usepackage[draft]{fixme}

\usepackage[capitalise]{cleveref}

\newcommand{\unity}{\ensuremath{\mathbb{1}}}
\newcommand{\ket}[1]{\ensuremath{| #1 \nobreak \rangle }}
\newcommand{\bra}[1]{\ensuremath{\langle #1 \nobreak | }}
\newcommand{\cnot}{\textsc{cnot}\xspace}
\newcommand{\swap}{\textsc{swap}\xspace}
\newcommand{\iswap}{$i$\swap}

\newcommand{\affA}{Department of Physics and Astronomy, Aarhus University, DK-8000 Aarhus C, Denmark}
\newcommand{\affB}{Aarhus Institute of Advanced Studies, Aarhus University, DK-8000 Aarhus C, Denmark}

\date{\today}
\begin{document}
	
	\title{Simple implementation of high fidelity controlled-$i$SWAP gates and quantum circuit exponentiation of non-Hermitian gates}
	
	\author{S. E. Rasmussen}
	\email{stig@phys.au.dk}
	\affiliation{\affA}
	\author{N. T. Zinner}
	\email{zinner@phys.au.dk}
	\affiliation{\affA}
	\affiliation{\affB}
	
	\begin{abstract}
		The \iswap gate is an entangling swapping gate where the qubits obtain a phase of $i$ if the state of the qubits is swapped. Here we present a simple implementation of the controlled-\iswap gate. The gate can be implemented with several controls and works by applying a single flux pulse. The gate time is independent of the number of controls, and we find high fidelities for any number of controls. We discuss an implementation of the gates using superconducting circuits and present a realistic implementation proposal, where we have taken decoherence noise and fabrication errors on the superconducting chip in to account, by Monte Carlo simulating possible errors. The general idea presented in this paper is, however, not limited to such implementations. An exponentiation of quantum gates is desired in some quantum information schemes and we therefore also present a quantum circuit for probabilistic exponentiating the \iswap gate and other non-Hermitian gates. 
	\end{abstract}
	
	\maketitle

	\section{Introduction}
	
	In order to perform non-trivial quantum computations it is often necessary to change the operation applied to one set of qubits depending upon the values of some other set of qubits. Some well known controlled gates are the controlled-\textsc{not} (\cnot), controlled-\textsc{swap} (\textsc{Fredkin}), and controlled-controlled-\textsc{not} (\textsc{Toffoli}) \cite{Nielsen2002}. While these gates are used in many quantum information schemes, such as quantum computing \cite{Vandersypen2001,Martin-Lopez2012,Lanyon2007}, error-correction \cite{Chuang1996,Barenco1997,Cory1998,Schindler2011}, cryptography \cite{Buhrman2001,Horn2005,Gottesman2001}, fault tolerant quantum computing \cite{Dennis2001,Paetznick2013}, and measurement \cite{Ekert2002,Fiuraifmmode2002}, they are not necessarily the most experimentally feasible ones \cite{Schuch2003}.
	
	Equivalent (in the sense that they both constitute a universal set of gates together with the set of one-qubit operations) to the \cnot gate is the \iswap gate which we denote $\hat S = \ket{00}\bra{00} \pm i (\ket{10}\bra{01} + \ket{01}\bra{10}) + \ket{11}\bra{11}$. The \iswap gate is a perfect entangling version of the \textsc{swap} gate, which is why it is equivalent to the \cnot gate. However, the \iswap gate has the advantage over the \cnot gate that it occurs naturally in systems with $XY$-interaction or Heisenberg models, such as solid state systems \cite{Tanamoto2008,Tanamoto2009}, superconducting circuits \cite{Zagoskin2006}, and in cavity mediated between spin qubits and superconducting qubits \cite{Imamoglu1999,Benito2019,Blais2004}. Other implementations of the \iswap gate include linear optics \cite{Wang2010,Bartkowiak2010} and nuclear spin using qudits \cite{Godfrin2018}.

	Despite several attemps of implementing the \iswap gate \cite{McKay2016,Dewes2012,Salathe2015}, the Fredkin gate \cite{Milburn1989,Chau1995,Fiuraifmmode2006,Fiuraifmmode2008,Gong2008,Patel2016,Ono2017,Smolin1996,Baekkegaard2019}, and other controlled-swapping gates \cite{Poletto2012,Rasmussen2019,Loft2020}, no one have embarked in the implementation of a \emph{controlled}-\iswap gate, to the best of our knowledge. Recently a deterministic Fredkin and exponential SWAP gate was implemented using three-dimensional, fixed frequency superconducting microwave cavities \cite{Gao2018,Gao2019}.
	
	Here we present a simple implementation of a multiqubit controlled-\iswap gate which we call \textsc{c}$^n$\iswap, where the $n$ indicates the number of control qubits. For a single control qubit this is essentially an $i$Fredkin gate, i.e., a Fredkin gate with a phase of $i$ on the swapping part.
	The implementation is based using the control qubits to tune the target qubits in and out of resonance by following the approach presented in Refs. \cite{Rasmussen2020,Christensen2020}, and can be realized using different schemes for quantum information processing. We include circuit design for an implementation of the \textsc{c}\iswap gate in superconducting circuits as well as for the \textsc{c}$^2$\iswap gate in the appendix. The gate requires a single flux pulse to operate, and the gate time is thus independent of the number of control qubits. When neglecting the decoherence of the qubits we find a fidelity above 0.998 for one control qubit. When including decoherence of the qubits the fidelity stays above 0.99 for up to four control qubits.
	
	Being able to exponentiate quantum gates can be useful in different quantum information schemes such as in continuous variable (CV) systems \cite{Braunstein2005}, where exponentiated gates, such as $\exp(i\theta \hat X)$, can be used to operate on the systems \cite{Lau2016,Lau2017}. Another scheme which might benefit from being able to exponentiate non-Hermitian quantum gates is quantum random walks \cite{Kempe2008}, where non-unitary operations is needed for, e.g., graph coloring \cite{Deervovic2018,Childs2003}. We therefore present a quantum circuit for probabilistic exponentiating of non-Hermitian operators, based on the method by \cite{Marvian2016} which works for exponentiating Hermitian operators. Our method is exact for cyclic operator, i.e., operators fulfilling $\hat T^n = \unity$, while it is approximate for all other non-Hermitian operators.
	
	The paper is organized as follows: In \cref{sec:implementation} we present a simple Hamiltonian and show how it yields an \textsc{c}$^n$\iswap gate. We discuss the effectiveness of the gate exploring the single qubit controlled-\iswap gate as an example in \cref{sec:Fredkin}. We further, in \cref{sec:superconducting}, present an implementation using superconducting circuits of the \textsc{c}\iswap gate and discuss how to expand it to more controls. In \cref{sec:SwappingArray} we show how to expand the implementation of the controlled-\iswap gate into controlling swapping of an array of qubits. In \cref{sec:simulations} we discuss a realistic implementation of the gate, including fabrication errors and decoherence noise.
	In \cref{sec:expGate} we present a quantum circuit for probabilistic exponentiating cyclic quantum gates, and discuss its range of validity. In \cref{sec:conclusion} we provide a summary
	and outlook for future work.

	\section{Implementation of the controlled-\lowercase{i}\textsc{swap} gate}\label{sec:implementation}
	 
	Consider $n+2$ qubits each with frequency $\omega_i$. The first $n$ qubits are connected to one of the two last qubits by an Ising couplings $J^z_i$, where $i$ refers to which of the first $n$ qubits. The last two qubits are further connected to each other by a transversal coupling $J^x$. We denote the last two qubits as target qubit $T1$ and $T2$. The Hamiltonian for the system is 
	\begin{equation}\label{eq:H}
	\begin{aligned}
	\hat H =& \hat H_0 - \frac{\Delta}{2}\sigma^z_{T1}  + \sum_{i=1}^{n} \frac{J_i^z}{2} \sigma^z_{T1}\sigma^z_i \\
	&+ \frac{J^x}{2} (\sigma^x_{T1} \sigma_{T2}^x + \sigma^y_{T1} \sigma_{T2}^y),
	\end{aligned}
	\end{equation}
	where $\sigma^{x,y,z}$ denotes the Pauli matrices, and the non-interacting part of the Hamiltonian is given as
	\begin{equation}\label{eq:H0}
	\begin{aligned}
	\hat H_0 =& - \sum_{i=1}^{n}  \frac{\omega_{i}}{2}\sigma^z_{i} - \frac{\omega_{T2}}{2}\left(\sigma^z_{T1}+\sigma^z_{T2}\right),
	\end{aligned}
	\end{equation}
	and $\Delta = \omega_{T1} - \omega_{T2}$ is the detuning of the two target qubits.
	Changing to the interaction picture using the transformation $\hat U_{\text{int}}(t) = \exp(i\hat H_0 t)$, the Hamiltonian takes the form
	\begin{equation}
	\begin{aligned}\label{eq:HI}
	\hat H_I =& - \frac{\Delta}{2}\sigma^z_{T1}  + \sum_{i=1}^{n} \frac{J_i^z}{2} \sigma^z_{T1}\sigma^z_i \\& + J^x(\sigma^+_{T1}\sigma^-_{T2} + \sigma^+_{T2}\sigma^-_{T1}).
	\end{aligned}
	\end{equation}
	In order to realize the behavior of the controlled \iswap gate we must require the detuning to be
	\begin{equation}\label{eq:condition}
	\Delta = -\sum_{i=1}^nJ_i^z,
	\end{equation}
	and $J_i^z \gg J^x$ for all $i$. Thus the energy shift due to the first $n$ qubits must be large enough to bring the last two qubit in and out of resonance, making the first $n$ qubits the control qubits and the last two the swapping qubits.
	 
	Changing into the frame rotating with the diagonal part of the Hamiltonian we obtain
	\begin{equation}
		\hat H_\text{rot} = J^x \left[ \sigma^+_{T1}\sigma^-_{T2} e^{i\sum_{i=1}^{n}J_i^z \left( 1 + \sigma^z_i \right)t} + \text{H.c.}\right].
	\end{equation}
	With the condition that $J^z \gg J^x$ both terms of $H_\text{rot}$ will rotate rapidly, and can thus be neglected using the rotating wave approximation, \emph{unless all} of the control qubits are in the state $|1\rangle$. The means that the Hamiltonian effectively becomes
	\begin{equation}
		\hat H_\text{rot} = J^x |\tilde 1 \rangle \langle \tilde 1 |_C \otimes \left[ \sigma^+_{T1}\sigma^-_{T2} +\sigma^+_{T2}\sigma^-_{T1}\right],
	\end{equation}
	where subscript $C$ denotes the state of the control qubits, i.e. the first $n$ qubits, and $T$ denotes the state of the target qubit, i.e., qubit $T1$ and $T2$. The state $|\tilde{1}\rangle_C = |11\dots 1\rangle_C$ denotes the state where all control qubits are in the state $|1\rangle$.
	 
	We can calculate the time evolution operator by taking the matrix exponential, $\hat U(t) = \exp(i\hat H_\text{rot}t)$, which yields
	\begin{equation}
	 	\begin{aligned}\label{eq:U}
	 	\hat U(t) =& \hat{\tilde{I}}_C \otimes \hat{I}_T + |\tilde{1}\rangle\langle \tilde{1}|_C\\
	 	&\otimes \begin{pmatrix}
	 	1 & 0 & 0 & 0\\
	 	0 & \cos(Jt)  &-i\sin(Jt)  & 0\\
	 	0 & -i\sin(Jt) & \cos(Jt)  & 0\\
	 	0 & 0 & 0 & 1	 \end{pmatrix},
	 	\end{aligned}
	\end{equation}
	where $\tilde{I}_C$ denotes the reduced identity of the control qubits where the states $|\tilde{1}\rangle\langle \tilde{1}|_C$ have been removed. The identity of the target qubits is denoted $\hat I_T$.
	 
	From \cref{eq:U} we see that for times $T = (2m+1)\pi/2J^x$, $m\in \mathbb{Z}$ the time evolution operator takes the form of a controlled-\iswap gate.
	\begin{equation}
	\hat U(t=T) = \hat{\tilde{I}}_C \otimes \hat I_T +|\tilde{1}\rangle\langle \tilde{1}|_C \otimes \hat S_T,
	\end{equation}
	where $\hat S_T$ is the two-qubit \iswap gate on the target qubits, which swaps the target qubit with a phase of $\pm i$. The phase on the target qubit depends on the sign of $\mp |J^x|$. For completeness we note that for times $T' = (2m+1)\pi/4J^x$ we obtain the controlled-$\sqrt{i\swap}$ gate \cite{Krantz2019}. Note that once time has passed such that the desired gate have been performed, interactions must be turned of. Thus the gate depends on control over the exchange interaction, which can be achieved differently depending on which scheme is used to implement the gate. In \cref{sec:superconducting} we present an implementation of the gate in superconducting circuits, where we also discuss how to control the exchange interaction.
	 
	\subsection{Example: The single controlled-\iswap gate}\label{sec:Fredkin}
	 
	In order to illuminate the performance of the system worked as a \textsc{c}$^n$\iswap gate we explore the example of the single controlled-\iswap gate. We chose this example since not only is it the simplest non-trivial example, it is also closely related to the Fredkin gate.
	A schematic presentation of the model yielding the controlled-\iswap gate can be seen in \cref{fig:fredkin}(a), which corresponds to \cref{eq:H} with $n=1$.
	 
	We characterize the performance of the gate by calculating the average process fidelity, which is defined as \cite{Nielsen2010,Nielsen2002,Horodecki1999,Schumacher1996}:
	\begin{equation}\label{eq:av_fidelity_formula}
	\bar{F} = \int d\psi\langle \psi|\hat U^\dagger\mathcal{E}(\psi)\hat U | \psi\rangle,
	\end{equation}
	where integration is performed over the subspace of all possible initial states and $\mathcal{E}$ is the quantum map realized by our system. We simulate the system using the Lindblad Master equation and the interaction Hamiltonian of \cref{eq:HI} using the QuTiP Python toolbox \cite{qutip}. The result is then transformed into the frame rotating with the diagonal of the Hamiltonian, and then the average fidelity is calculated.
	 
	\begin{figure}
		\centering
	 	\includegraphics[width=\columnwidth]{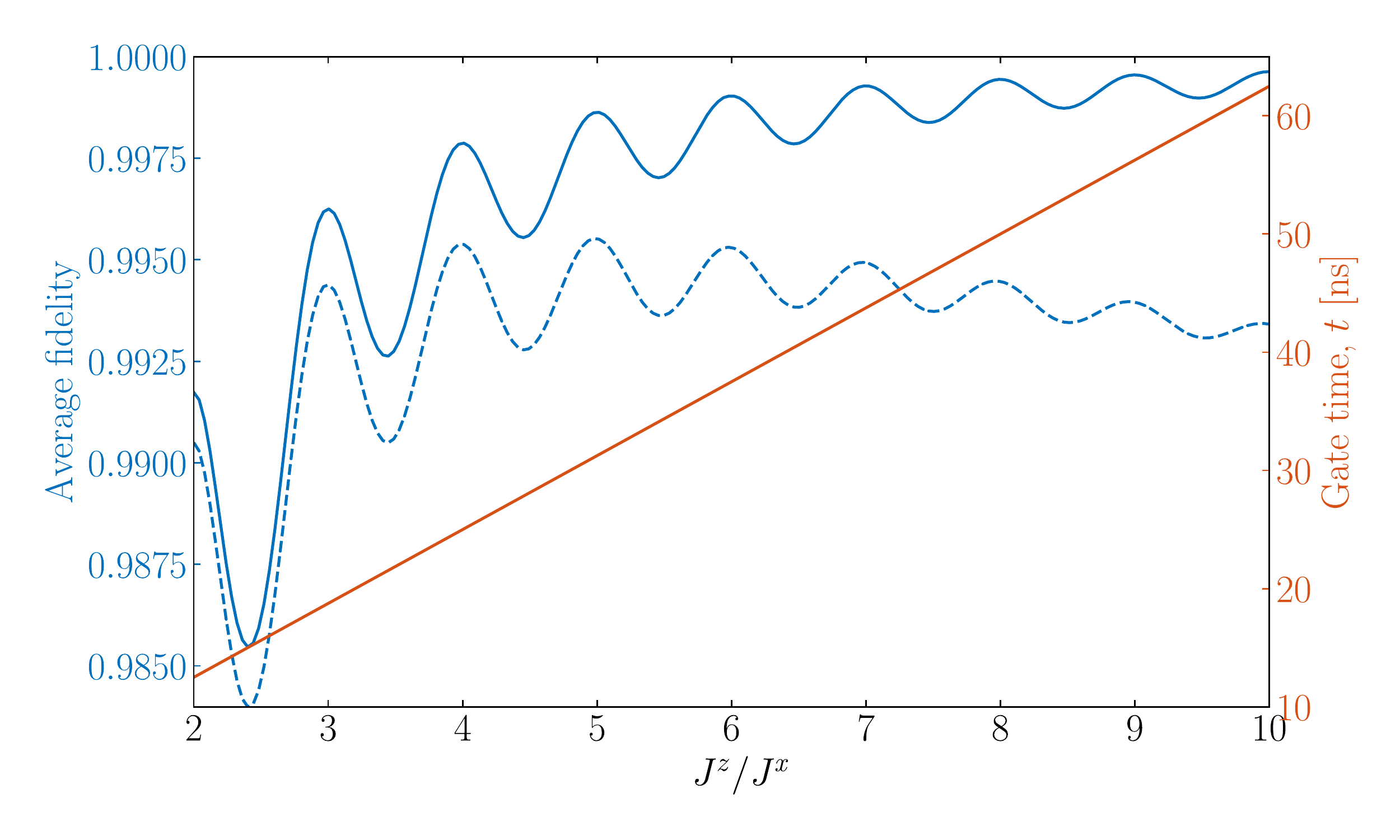}
	 	\caption{Simulation of the controlled-\iswap gate for different values of the coupling $J^x$. The blue lines indicates the average fidelity (left $y$-axis), while the straight red line indicates the gate time $T$ (right $y$-axis). The dashed blue line is the average fidelity with a decoherence time of $T_1 = T_2 =\SI{30}{\micro\s}$, while the solid is without decoherence.}
	 	\label{fig:simulation}
	\end{figure}
 
	For all simulations we have $J^z/2\pi = \SI{50}{\MHz}$, while we change the transversal coupling, $J^x/2\pi$, from 5 to \SI{25}{\MHz}. The average fidelity of the simulation can be seen in \cref{fig:simulation} together with the gate time. The figure shows both the average fidelity without any decoherence and with a decoherence time of  $T_1 = T_2 =\SI{30}{\micro\s}$ \cite{Wendin2017}. We model decoherence as relaxation and phase errors, we do not include excitation by thermal photon, as it contributes very little to the decoherence \cite{Jin2015}. Without any decoherence we find that the average fidelity increases asymptotically towards unity as the driving decreases, with the only expense being an increase in gate time. Since decoherence increases over time, a longer gate time means lower fidelity, which is exactly what we observe when including decoherence in the simulations. In this case we find that the fidelity peaks at $\sim0.995$ around $J^z/J^x \sim 4$, which yields a gate time of $T\sim \SI{25}{\nano\s}$. However, we note that the fidelity are dependent on the parameters $J^x$ and $J^z$ thus changing these will change the fidelity. We also see that for just $J^z = 2 J^x$ we obtain an average fidelity above 0.99 for a gate time $T\sim \SI{15}{\nano\s}$. 
	The oscillation of the average fidelity is due to a small mismatch in the phase of the evolved state compared to the desired matrix in \cref{eq:U}, which disappears when $J^z/J \in \mathbb{Z}$.
	
\begin{figure}
		\centering
		\includegraphics[width=\columnwidth]{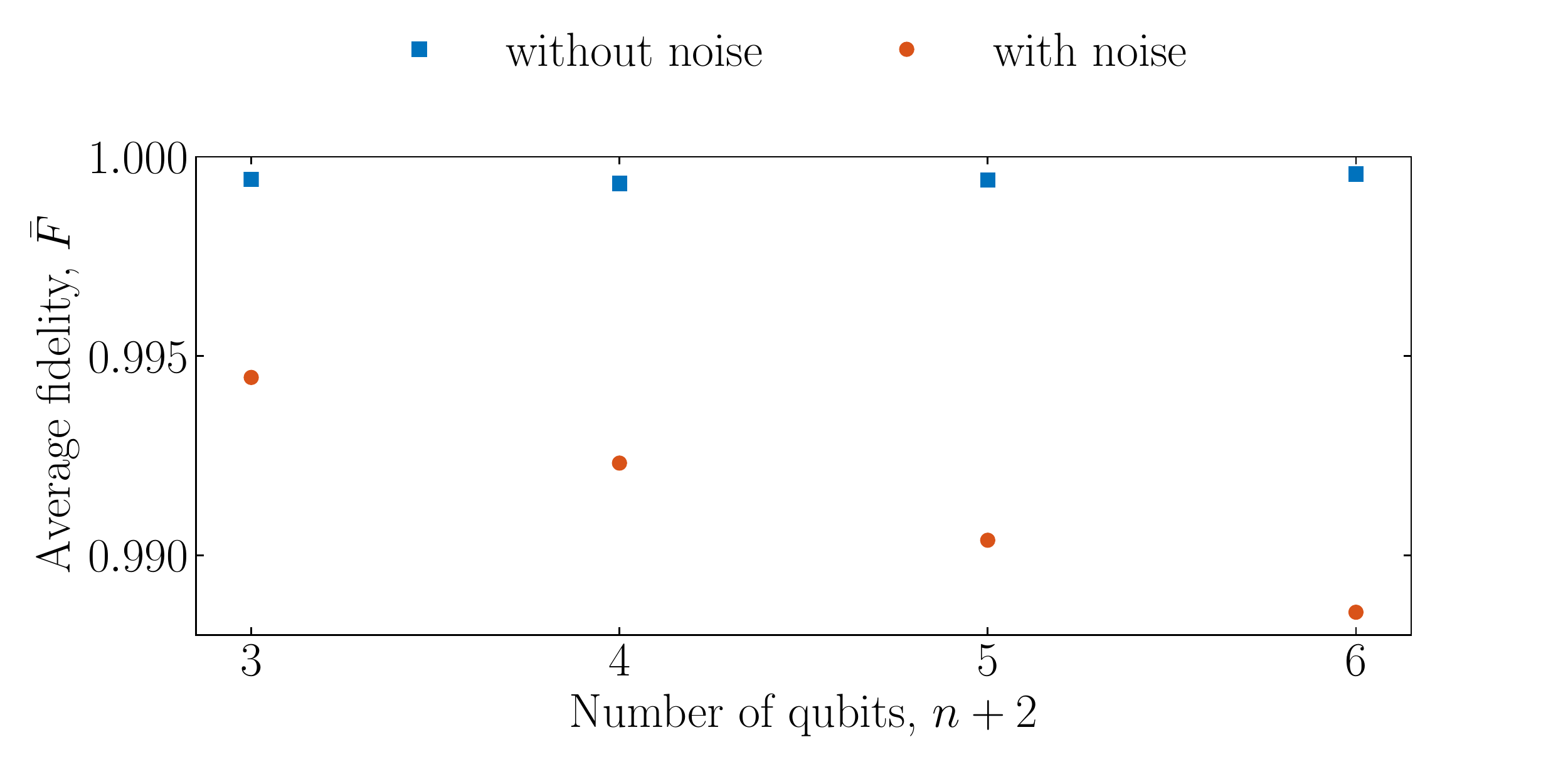}
		\caption{Average fidelity as a function of the number of qubits in the C$^ni$\textsc{swap} gate. The blue square markers indicate the simulation without decoherence, while the round red markers indicates the simulation done with a decoherence time of $T_1 = T_2 =\SI{30}{\micro\s}$. All simulations are done with $J^z/J=5$, i.e., peak fidelity cf. \cref{fig:simulation}.}
		\label{fig:control}
	\end{figure}

	We simulate the C$^ni$\textsc{swap} gate for different $n$ in the optimal ratio between couplings, $J^z/J^x \sim 4$. The result of this simulation is seen in \cref{fig:control}. We observe that the fidelity stays above 0.998 for up to $n=4$ control qubits when decoherence is not included. The reason for this is that for larger $n$ the gate resembles the identity more. This is due to the fact that the identity operation is applied to the control qubits, meaning that for a large number of control qubits, the gate will perform the identity on the control qubits and the swapping operation will only be performed on the target qubits. When decoherence is included the average fidelity decreases for larger $n$ as it should, however, we still find a fidelity above 0.99 for up to 4 controls.

	\section{Experimental implementation in superconducting circuits}\label{sec:superconducting}

	\begin{figure}
	 	\centering
	 	\includegraphics[width=\columnwidth]{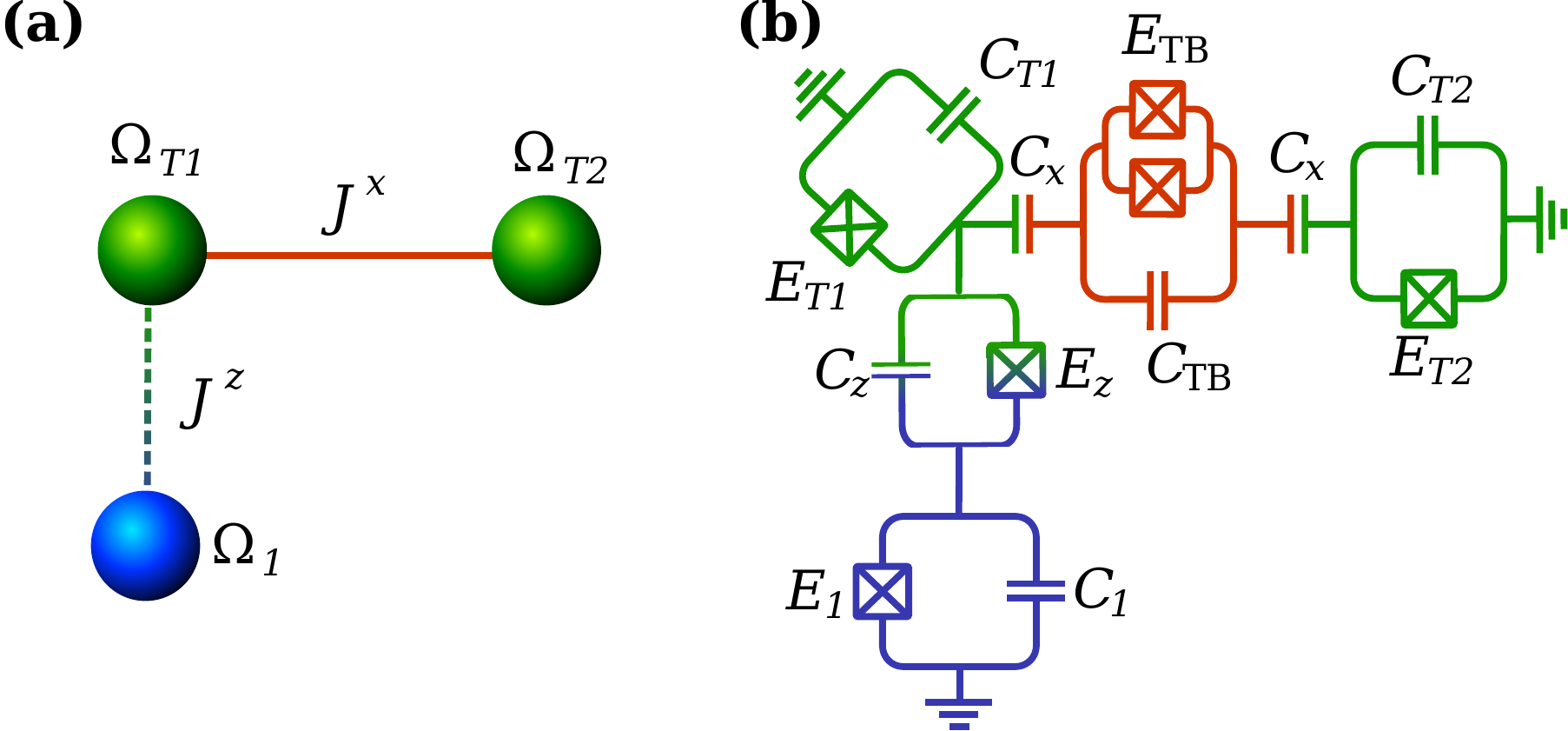}
	 	\caption{Implementation of the controlled-\iswap gate, i.e., iFredkin gate. Figure \textbf{(a)} shows a schematic representation of the model implementing the controlled-\iswap gate, with the green spheres (subscript $T1$ and $T2$) representing the target qubits and the blue sphere (subscript 1) representing the control qubit. The red line indicates an interaction which can be turn on and off, how to do this depends on the qubit implementation, see\cref{sec:superconducting} for an example of how to do this. Figure \textbf{(b)} shows the superconducting circuit yielding the model in figure \textbf{(a)}. The different parts of the system are colored according to their role, as per \textbf{(a)}.}
	 	\label{fig:fredkin}
	\end{figure}
 
 	A possible implementation of the controlled-\iswap gate using superconducting circuits can be seen in \cref{fig:fredkin}(b). The circuit consists of three fixed frequency transmon qubits \cite{Koch2007,Schreier2008}, where two of them are connected through a tunable bus qubit, following the approach by Ref. \cite{McKay2016}, and the third qubit is connected to the other two by Josephson junctions, with as small a parasitic capacitance as possible.
 	
 	After eliminating the superficial degree of freedom of the tunable bus the Hamitonian of the circuit takes the form
 	\begin{align}\label{eq:HCircuit}
 	\hat H =& \frac{1}{2} \hat{\vec{p}}^T K^{-1} \hat{\vec{p}} - \sum_{\substack{i=\{1,T1,\\T2,TB\}}} E_i\cos\hat \varphi_i
 	- E_z\cos(\hat\varphi_{T1}-\hat\varphi_1)\nonumber \\ &- 2 E_{TB} \cos(2\Phi) \cos (2\varphi_{TB}) ,
 	\end{align}
 	where $\varphi_i$ are the node fluxes, $\vec{p}^T = (p_{T1},p_{T2},p_1)$ are the conjugate momenta, $\Phi$ is the external flux through the tunable bus, and $K$ is the capacitance matrix. 
 	
 	As the capacitive couplings yields transversal $XX$-couplings when truncating to a Ising-type model, we are not interested in the capacitive couplings between the control qubit and the target qubits, and thus we require $C_z \ll C_i, C_{Ti}$ which will leave the capacitance matrix being approximately diagonal, with the exception of the desired capacitance between the target qubits and the tunable bus. This leaves only longitudinal $ZZ$-couplings between the control and target qubit. This limit where the longitudinal coupling dominates over the transversal couplings is within experimental reach \cite{Kounalakis2018}. An other way of reaching high-contrast $ZZ$-couplings could be to use a combination of transmon and flux qubit, and then engineering opposite sign anharmonicities as in Ref. \cite{Zhao2020}. When truncating the Hamiltonian in \cref{eq:HCircuit} to a two level system, we follow the approach presented by Ref. \cite{McKay2016} and adiabatically remove the tunable bus qubit, by considering the dispersive regime $|g^x_{TjTB}/(\omega_{Tj} - \omega_{TB})| \ll 1$, which yields the following Hamiltonian
 	\begin{equation}
 	\begin{aligned}
 	\hat H =& \sum_{i=1}^{n} \frac{\omega_i}{2}\sigma^z_i + \frac{\tilde \omega_{T1}(\Phi)}{2}\sigma^z_{T1} + \frac{\tilde \omega_{T2}(\Phi)}{2}\sigma^z_{T2} \\
	&  + \frac{J_{1}^z}{2}\sigma^z_{T1}\sigma_1^z + \tilde J^x(\Phi)\left(\sigma^+_{T1}\sigma^-_{T2} + \sigma^-_{T1}\sigma^+_{T2}\right),
 	\end{aligned}
 	\end{equation} 
 	where the tildes indicates dressed qubit frequency and coupling stemming from the removal of the tunable bus qubits.
 	 	
 	Now by applying the external flux as a sinusoidal fast-flux bias modulation such that the external flux is $\Phi(t) = \Theta + \chi \cos (\omega_\Phi t)$ we can time average over the qubit frequencies and the exchange coupling, and thus gain control over these parameters. By turning on and off the sinusodial part of the flux pulse we can turn the gate on and off as well. After time averaging the Hamiltonian takes the form
 	\begin{equation}
 	\begin{aligned}
 	\hat H =& \sum_{i=1}^{n} \frac{\omega_i}{2} \sigma^z_i + \frac{\bar \omega_{T2}}{2} (\sigma^z_{T1} + \sigma^z_{T2})  -\frac{\bar\Delta_{T1}}{2}\sigma_{T1}^z \\ &+ \sum_{i=1}^n \frac{J_{i}^z}{2}\sigma^z_{T1}\sigma_i^z + \bar J^x\left(\sigma^+_{T1}\sigma^-_{T2} + \sigma^-_{T1}\sigma^+_{T2}\right),
 	\end{aligned}
 	\end{equation}
 	where the bar indicates time average and $\bar \Delta_{T1} = \bar\omega_{T1} - \bar\omega_{T2}$. The time averaged of the exchange coupling is, to second order,
 	\begin{equation}
 		\begin{aligned}
 		\bar J^x(\Phi(t), t)  = & \left. \frac{\partial \tilde J^x}{\partial \Phi} \right|_{\Phi \rightarrow \Theta} \chi \cos (\omega_\Phi t)\\
 		&+ \frac{\chi^2}{4} \left. \frac{\partial^2 \tilde J^x}{\partial \Phi^2} \right|_{\Phi \rightarrow \Theta} \cos (2\omega_\Phi t),
 		\end{aligned}
 	\end{equation}
 	where we note that the coupling depends both on the external flux, but also on explicitly on the time, which means that the coupling strength will oscillate in time. Changing into a frame rotating with the diagonal of the Hamiltonian, we find
 	\begin{equation}\label{eq:barJx1}
 		\hat H = \bar J^x(\Phi,t) e^{i(\bar \Delta_{T1} -  J^z_i\sigma_i^z) t}\left(\sigma^+_{T1}\sigma^-_{T2} + \sigma^-_{T1}\sigma^+_{T2}\right).
 	\end{equation}
 	If we require the frequency of the alternating part of the external flux to be resonant with the phase of the Hamiltonian when the control qubit is in the state $|1\rangle$, i.e. 
 	\begin{equation}\label{eq:omegaPhi}
 		\omega_\Phi = \bar \Delta_{T1} + J^z_i,
 	\end{equation}
 	we can use the rotating wave approximation, and remove all terms, except when the control qubit is in the $|1\rangle$ state. Thus in condition in the original implementation, \cref{eq:condition} is now replaced by the more easily obtainable expression in \cref{eq:omegaPhi}. This also mean that the gate time becomes $T=(2m+1)\pi/J^x$ due to the cosine function. Nevertheless the result is the same and we obtain a controlled \iswap gate.
 	
 	Note that there is also a resonant coupling at $ 2\omega_\Phi = \bar \Delta_{T1} + J^z_i$ in which case the exchange coupling is via the second order term in \cref{eq:barJx1}. This could be used to lower the coupling in order to satisfy the requirement $J^x \ll J^z$.
 	
 	A detailed calculation going from the circuit design to the gate Hamiltonian can be found in \cref{app:analCircuit} together with an example of an implementation of the \textsc{c}$^2$\iswap gate.
	An alternative approach to implementing such a tunable exchange coupling is to use the "gmon"-based design proposed in Ref. \cite{Chen2014}.
	
\subsection{Simulations}\label{sec:simulations}

\begin{figure}
	\centering
	\includegraphics[width=\columnwidth]{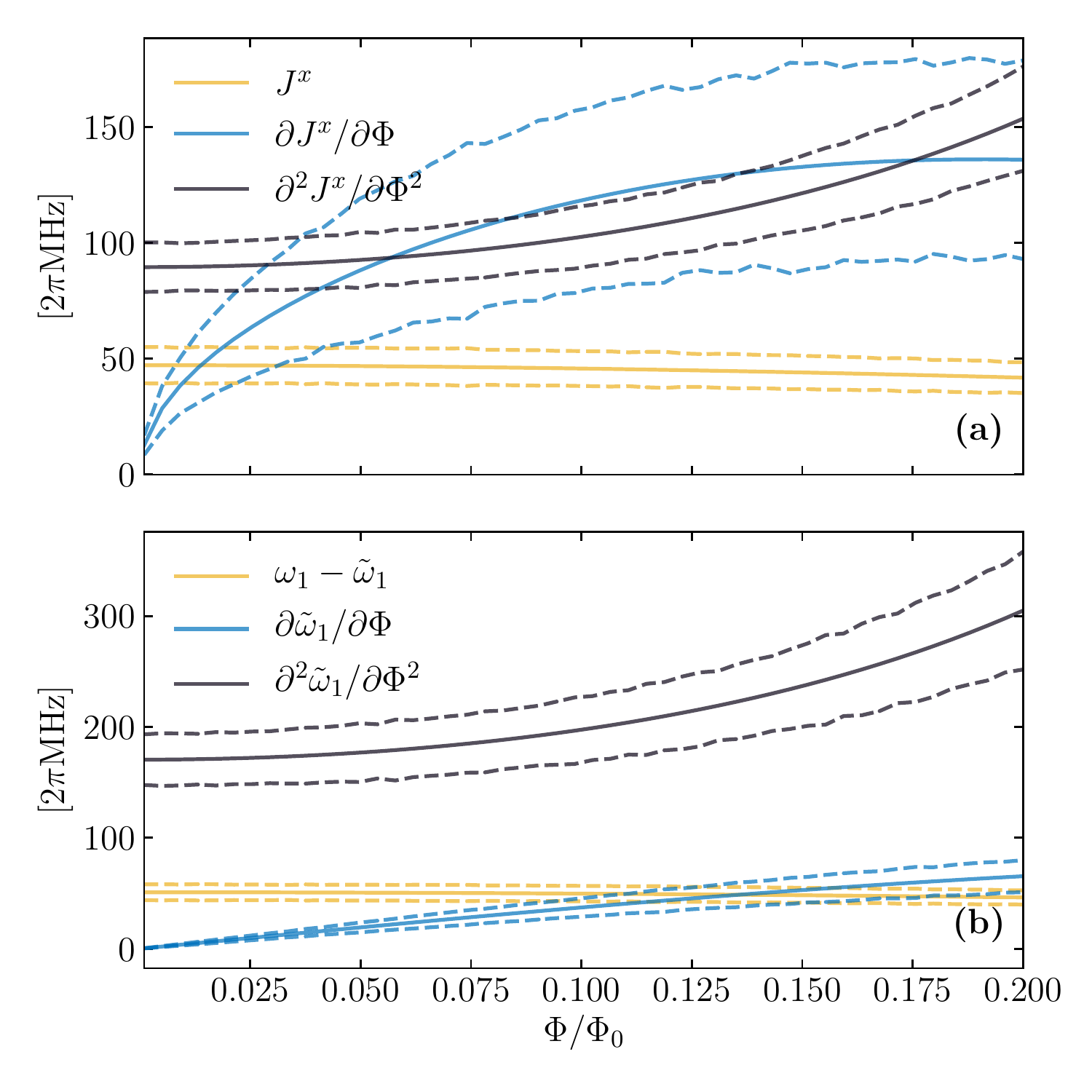}
	\caption{Typical derivatives of the gate parameters $J^x$ \textbf{(a)} and $\tilde\omega$ \textbf{(b)}. The dashed lines indicates the error on the parameters, found using Monte Carlo simulations. In particular the parameters comes from column 2 in \cref{tab:gateParams}.}
	\label{fig:JxvsPhi}
\end{figure}

In order to show that the superconducting circuit model presented in the previous section does indeed give the desired result, we find realistic parameters for the circuit presented in \cref{fig:fredkin} and their corresponding gate parameters. These parameters can be found in \cref{tab:circuitParams,tab:gateParams,tab:qualiParams} for $\Phi=0$. In \cref{fig:JxvsPhi} we present typical parameters relevant for the gate implementation, i.e., derivatives of $J^x$ and $\tilde\omega$ as a function of the external flux, $\Phi$. In a realistic implementation the circuit parameter are not perfect compared to the ones found in our simulations. Therefore we simulate with errors. We assume a fabrication error of up to 10\% of 95\% of the simulations. We then Monte Carlo simulates the circuit in order to find the error on the gate parameters. These errors are presented as the dashed lines in \cref{fig:JxvsPhi}. While these error might seem large they are not a problem for the gate, as the gate operation is mainly dependent on \cref{eq:omegaPhi}, which can be achieved only with control over just the external flux.

\begin{figure}
	\centering
	\includegraphics[width=\columnwidth]{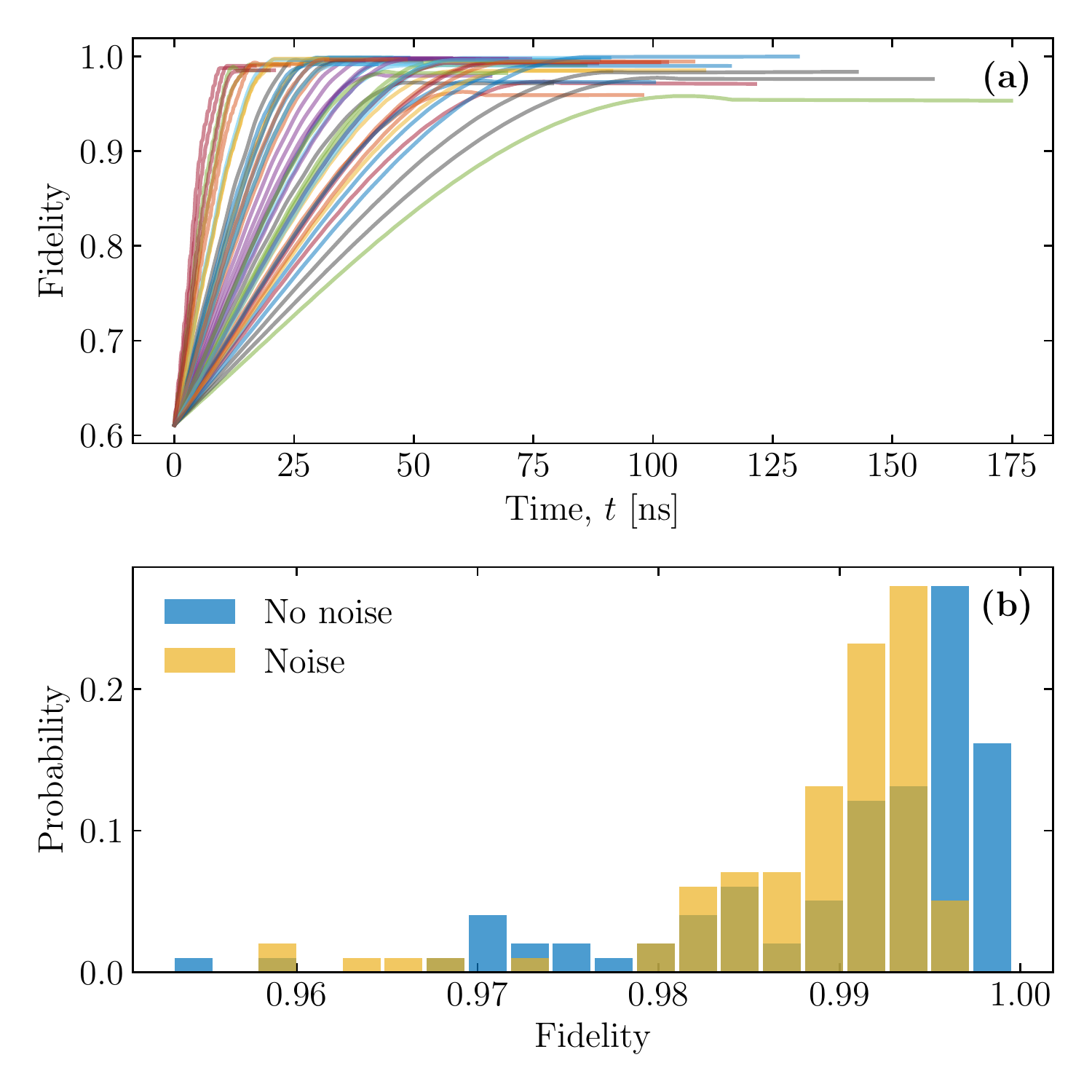}
	\caption{\textbf{(a)} Average fidelity of the Monte Carlo simulation of the gate, as a function of time. All shown simulation are without noise \textbf{(b)} Distribution of fidelities of the simulations at the gate time. The noise is the same as in \cref{sec:Fredkin}.}
	\label{fig:fidelity}
\end{figure}

Using the gate parameters found in \cref{tab:gateParams,fig:JxvsPhi} we simulate the gate using an external DC flux of $\Phi = 0.100 \Phi_0$ and a modulation of $\chi= 0.100\Phi_0$. The external flux frequency is determined from \cref{eq:omegaPhi}, however we include an error corresponding to a standard deviation of $1\si{\MHz}/2\pi$ in our simulation. In Ref. \cite{McKay2016} they have an error of $0.1\si{\MHz}/2\pi$. The result of these Monte Carlo simulations can be seen in \cref{fig:fidelity} where we have plotted the average fidelity of a subset of the simulations as a function of time. From the distribution of the fidelities we see that 60\% of the simulations end up with a fidelity above 0.99, while 90\% of the simulations are above 0.98 when the simulation is done without decoherence noise, while the fidelity is smaller when decoherence noise is included in the simulations.

We conclude that even when including significant errors in the fabrication of the circuit, the gate still yields a high fidelity with the controlled-\iswap gate.

\section{Controlled swapping arrays}\label{sec:SwappingArray}

Suppose we have multiple qubits which we want to swap in a controlled way, i.e., first swapping two qubits, then swapping two other qubits, and so on. This might be useful in a range of quantum algorithms.

In this section we discuss how to expand the idea of the controlled-\iswap gate previous section into a system where we can swap qubits in an array arbitrarily. We will discuss this for the case of an array of first three qubits and then briefly for four qubits, but the ideas will be easily expandable to more qubits. 

\begin{figure}
	\centering
	\includegraphics[width=\columnwidth]{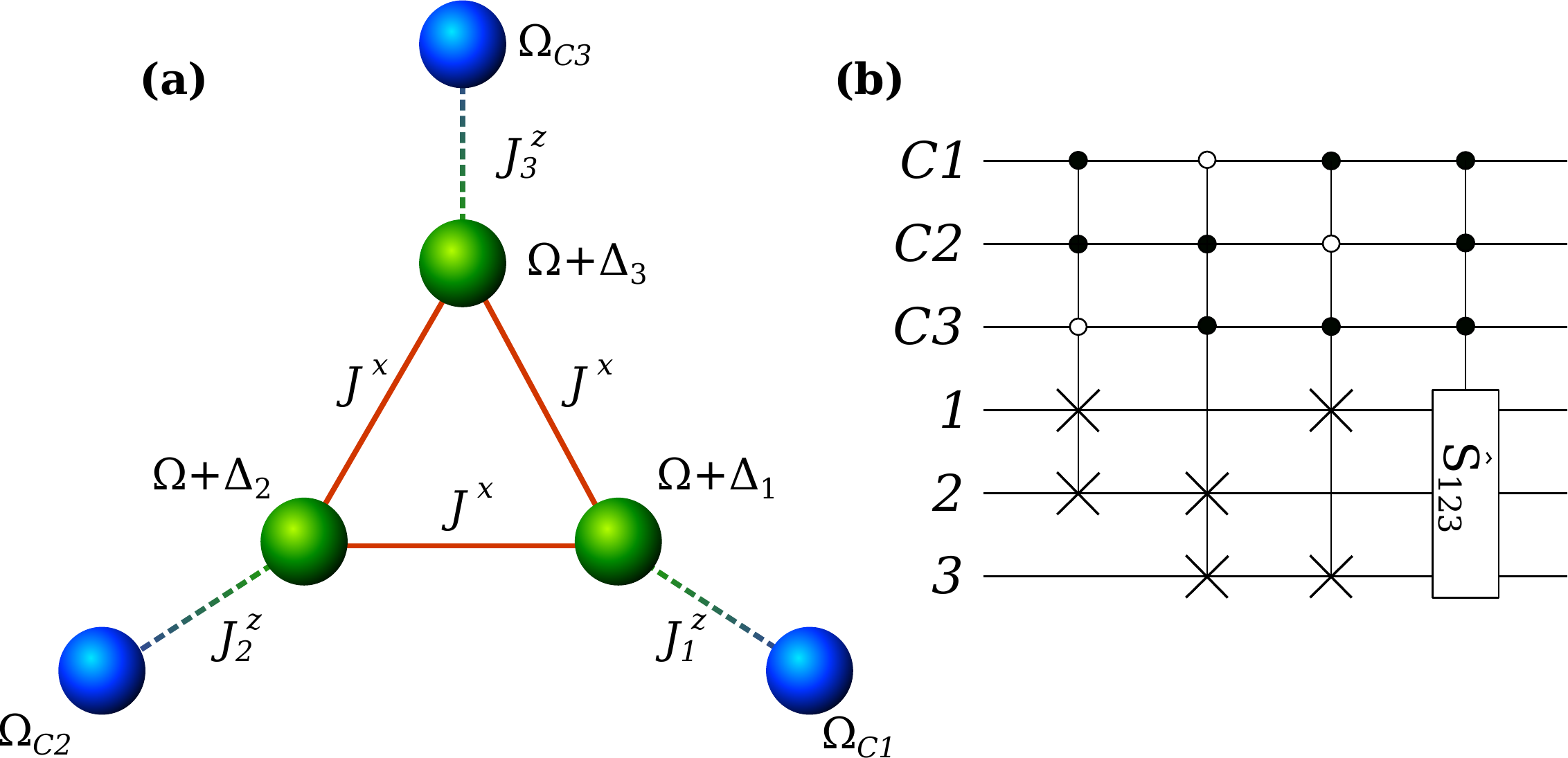}
	\caption{\textbf{(a)} Schematic representation of the model leading to controlled swapping between three qubits. The green spheres (subscripts 1, 2, and 3) represent the swapping qubits, while the blue spheres (subscripts $C1$, $C2$, and $C3$) represent the ancilla qubits, which controls the swapping. \textbf{(b)} Quantum circuit representations of the model in \textbf{(a)} respectively for times $T=(2m+1)\pi/(2J^x)$, $m\in \mathbb{Z}$. The top three ancilla qubits controls the swapping and corresponds to the blue spheres, while the lower three qubits corresponds to the green spheres. The filled circles indicates that the ancilla qubits must be the the state $|1\rangle$ for the swap to be activated, while the non-filled circles indicates that the ancilla qubits must be in the state $|1\rangle$; This corresponds to the time evolution operators in \cref{eq:SwappingArray3}.}
	\label{fig:SwappingArray3}
\end{figure}

In an attempt to create such a system we connect all qubits which we wish to be able to swap to each other with transversal coupling, $J^x$, each of these $n$ qubits are detuned from the average frequency of the qubits, such that $\Delta_i \neq \Delta_j$ for $i\neq j=1,2,\dots ,n$. Following the idea of \cref{fig:fredkin}(a) we add a control qubit for each target qubit, and couple it with Ising couplings, $J^z_i$ to each qubit. A schematic representation of the model for $n=3$ can be seen in \cref{fig:SwappingArray3}(a). The Hamiltonian for such a system becomes
\begin{equation}\label{eq:SwappingArray3H}
\begin{aligned}
\hat H =& - \sum_{i=1}^{n} \left[\frac{\omega + \Delta_i}{2}\sigma^z_{Ti} + \frac{\omega_{Ci}}{2}\sigma_{Ci}^z \right] \\ & + \sum_{ i=1}^{n} \frac{J^z_i}{2}\sigma^z_{Ti} \sigma^z_{Ci} + \frac{1}{4} \sum_{j\neq i=1}^{n} J^x  \sigma_{Ti}^x \sigma_{Tj}^x.
\end{aligned}
\end{equation}
where $\omega$ is the average over all the target qubits frequency, $\Delta_i$ is the detuning of the $i$th target qubit from the average frequency of the target qubits, and the subscript $Ti$ indicates the $i$th target qubit, while the subscript $Ci$ indicates the $i$th control qubit.

If we require that the Ising couplings have the strengths $J_i^z = -\Delta_i$, and require that $J^z_i \gg J^x$ for all $i$, then at times $T = (2m+1)\pi/(2J^x)$, $m\in \mathbb{Z}$ the time evolution operator for the $n=3$ case takes the form
\begin{equation}\label{eq:SwappingArray3}
\begin{aligned}
\hat U(T) =& \hat{\tilde{I}}_C \otimes \hat I_T +  |110\rangle\langle 110|_C \otimes \hat S_{12} \\ &+|011\rangle\langle 011|_C \otimes \hat S_{23} +|101\rangle\langle 101|_C \otimes \hat S_{13}\\ & + |111\rangle\langle 111|_C \otimes \hat S_{123}.
\end{aligned}
\end{equation}
where $\hat{\tilde{I}}_C$ denotes the reduced identity of the control qubits where the states $|100\rangle\langle 100|_C, |010\rangle\langle 010|_C$, and $|001\rangle\langle 001|_C$ have been removed. The identity of the three target qubits is denoted $\hat{I}_T$, and $\hat S_{ij}$ is the two-qubit \iswap gate which swaps the state of the qubits $i$ and $j$. The quantum circuit of the model can be seen in \cref{fig:SwappingArray3}(b). 

From the time evolution operator in \cref{eq:SwappingArray3} we see that we have complete control over which qubits we wish to swap, depending on the three ancilla qubits, i.e., if we wish to swap qubits $Ci$ and $Cj$ to be in the $\ket{1}$ state and remaining control qubits to be in the state $\ket{0}$, in which case with the $\pm $\iswap-operators $\hat S_{ij}$ swaps the state of the two qubits $i$ and $j$. We note that we also obtain a three-way swapping operator when all control qubits are in the $|1\rangle $ state. In its matrix representation the three-way swap-operator is an $8\times 8$ matrix and takes the form
\begin{equation}
\hat S_{123} = \begin{pmatrix}
1 & 0 & 0 & 0 \\
0 & \hat S_{1} & 0 & 0 \\
0 & 0 & \hat S_{2} & 0 \\
0 & 0 & 0 & 1
\end{pmatrix},
\end{equation}
where the two operators $\hat S_1$ and $\hat S_2$ are $3\times 3$ matrices and operate on the three dimensional subspaces of one and two excitation number, of the target subspace, respectively. In their matrix representation these take the same form
\begin{widetext}
\begin{equation}\label{eq:Sijk}
\hat S_{1,2} = \frac{1}{3}e^{iJ^xt/2}\begin{pmatrix}
3\cos (3J^xt/2) - i\sin (3J^xt/2) & 2i \sin(3J^xt/2) & 2i \sin(3J^xt/2) \\
2i \sin(3J^xt/2) & 3\cos (3J^xt/2) - i\sin (3J^xt_/2) & 2i \sin(3J^xt/2) \\
2i \sin(3J^xt/2) & 2i \sin(3J^xt/2) & 3\cos (3J^xt/2) - i\sin (3J^xt/2)
\end{pmatrix},
\end{equation}
\end{widetext}
which can be used to entangle all three qubits. We consider the special case of $T' = m\pi/3J^x$, $m\in \mathbb{Z}$, for which the operator takes the form
\begin{equation}
\hat S_{1,2} = \frac{1}{3}ie^{i\pi/6}\begin{pmatrix}
-1 & 2  & 2  \\
2  & -1 & 2  \\
2  & 2  & -1
\end{pmatrix}.
\end{equation}
This operator can be used to create state belonging to the same non-biseparable classes of three-qubit states as the $W$ state \cite{Dur2000}.

\begin{figure}
	\centering
	\includegraphics[width=0.9\columnwidth]{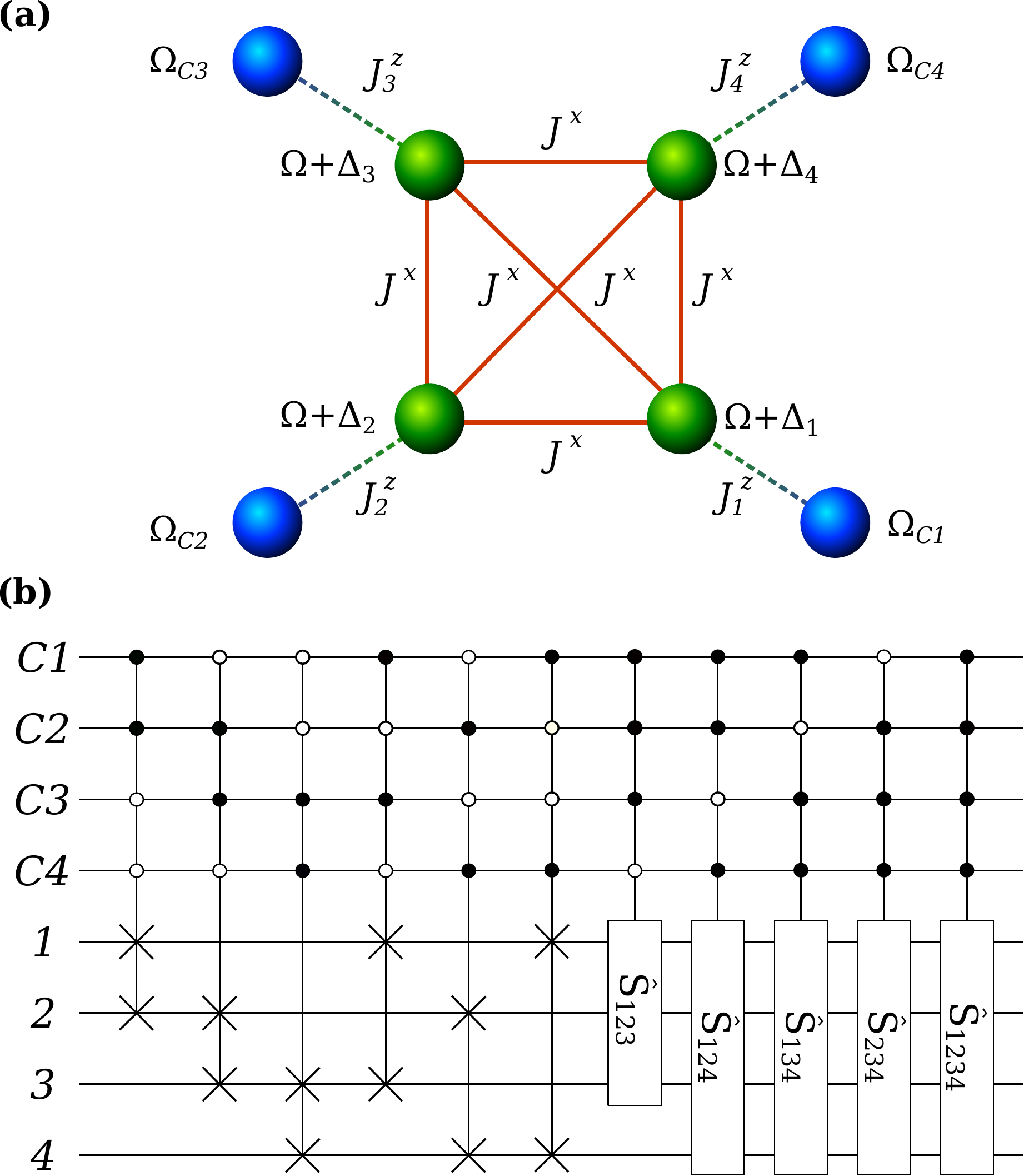}
	\caption{\textbf{(a)} Schematic representation of the model leading to controlled swapping between four qubits. The green spheres (subscript 1, 2, 3, and 4) represent the swapping qubits, while the blue spheres (subscript $C1$, $C2$, $C3$, and $C4$) represent the ancilla qubits, which controls the swapping. \textbf{(b)} Quantum circuit representations of the model in \textbf{(a)} for times $T=(2m+1)\pi/(2J^x)$, $m\in \mathbb{Z}$. The top four ancilla qubits controls the swapping and corresponds to the blue spheres, while the lower four qubits corresponds to the green spheres.}
	\label{fig:SwappingArray4}
\end{figure}

In \cref{fig:SwappingArray4}(a) we show the model for a four qubit swapping array with all-to-all couplings corresponding to Hamiltonian in \cref{eq:SwappingArray3H} with $n=4$. In \cref{fig:SwappingArray4}(b) we present the corresponding gate of the model coming from making the time evolution operator from the Hamiltonian. As above we obtain fully controllable two-qubit swapping between all of the four qubits. We further obtain four three-qubit entangling gates, similar to the one in \cref{eq:Sijk} and one single four-qubit entangling gate.

In order to test the viability of our analysis we simulate the Hamiltonian in \cref{eq:SwappingArray3H} using the Python toolbox QuTiP using the same approach as in \cref{sec:Fredkin}. Using parameters $J^z_i/(2\pi) \in \{-20,20,60\} \si{\MHz}$ and $J^x = \min_i |J^z_i|/5$ we find a fidelity of 0.993 at time $T = \pi/(2J^x) = \SI{62.5}{\nano\s}$ without including decoherence, and a fidelity of 0.98 when including a decoherence time of $T_1 = T_2 = \SI{30}{\micro\s}$.

\section{Probabilistic exponentiating of cyclic non-Hermitian quantum gates}\label{sec:expGate}

In this section we present an exact probabilistic method for exponentiating cyclic non-Hermitian gates using an explicit quantum circuit. While our method is exact for cyclic operators it is approximate for non-cyclic operators.
The controlled-\iswap gate presented in this paper is in fact a cyclic non-Hermitian gate. Note that exponentiating non-Hermitian gates leads to non-unitary gates.

Unitary Hermitian gates can be exponentiated using the method developed by Marvian and Lloyd \cite{Marvian2016}. Albeit they only present their method for the controlled-\textsc{swap} gate, it works for all unitary Hermitian gates. Here we extend their method in order to exponentiate non-Hermitian gates. Our method is exact for a gate, $\hat T$, for which $\hat T^n = \unity$ for $n\in\mathbb{Z}$ and approximately correct if this is not the case. We call gates where $\hat T^n = \unity$ for cyclic gates with cyclic order $n$. For $n>2$ all cyclic gates become non-Hermitian, due to the fact that all eigenvalues of Hermitian matrices must be real and a diagonal matrix $D$ fulfilling the Spectral theorem such that $\hat T =\hat U\hat D\hat U^{-1}$, where $\hat U$ is a unitary, must then fulfill $\hat D^n = \unity$.

Our result become interesting as soon as you want to exponentiate some sort of phase gate, with a phase other than $-1$, in which case the gate becomes non-Hermitian. This means that the result of such exponentiating will be non-unitary for $n>2$. In \cref{tab:non-HermitianGates} we mention a few often used non-Hermitian gates and their cyclic order. We note that in order to use our method we must be able to perform a controlled version of the gate we wish to exponentiate, i.e., if we wish to exponentiate an \iswap we would need a controlled-\iswap, as discussed above.

\begin{table}
	\centering
	\caption{Common non-Hermitian quantum gates and their cyclic order $n$. Note that we assume $\phi$ to be $\pi$ divided by an integer. The controlled version of the gates mentioned in this table are also non-Hermitian with the same cyclic order.}
	\label{tab:non-HermitianGates}
	\begin{tabular}{llc}
		\hline
		Gate & & $n$ \\
		\toprule
		Phase shift & $R_\phi$ & $\pi/\phi$ \\
		Square root of not &$\sqrt{\textsc{not}}$ & 4 \\
		Imaginary swap & \iswap & 4 \\
		Square root of swap &$\sqrt{\text{\textsc{swap}}}$ & 4 \\
		Ising $XX$ coupling & $XX_0$ or $XX_{\pi}$ & 8 \\
		Ising $YY$ coupling & $YY_{\phi}$ & $2\pi/\phi$ \\
		Ising $ZZ$ coupling & $ZZ_{\phi}$ & $2\pi/\phi$ \\
		Deutch & $D_\phi$ & $2\pi/\phi$ \\
		\hline
	\end{tabular}
\end{table}

\begin{figure}
	\begin{equation*} \Qcircuit @C=1em @R=0.5em @!R  {
			\lstick{}  & \ctrl{5} & \qw & \qw & \qw & \qw & \meter\\
			\lstick{}  & \qw & \ctrl{4} & \qw & \qw & \qw & \meter\\
			\lstick{}  &  &   & \vdots  &  & \\
			\lstick{}  & \qw & \qw & \qw & \ctrl{2} & \qw & \meter\\
			\lstick{}  & \qw & \qw & \qw & \qw & \ctrl{1} & \meter \inputgroupv{1}{5}{.8em}{3.7em}{\ket{\tilde{\varphi}}}\\
			\lstick{}  & \multigate{1}{\hat T} & \multigate{1}{\hat T} & \qw & \multigate{1}{\hat T}& \multigate{1}{\hat T} &\qw\\
			\lstick{}  & \ghost{T} & \ghost{T} & \qw & \ghost{T} & \ghost{T} &\qw  \inputgroupv{6}{7}{.8em}{0.9em}{\ket{\gamma}}} \end{equation*}
	\caption{Quantum circuit used to exponentiate a matrix $\hat T$ for which $\hat T^n=\unity$. On top we have $n-1$ ancilla qubits which are prepared in the state $\ket{\tilde{\varphi}}$. Each acts as a conditional for a $\hat T$ operation, and finally they are all measured in the $\{\ket{\pm}\}$-basis. Note that the $\hat T$ operation does not have to be a two qubit operation, it can operate on $m$ qubits.}
	\label{fig:qcircuit}
\end{figure}
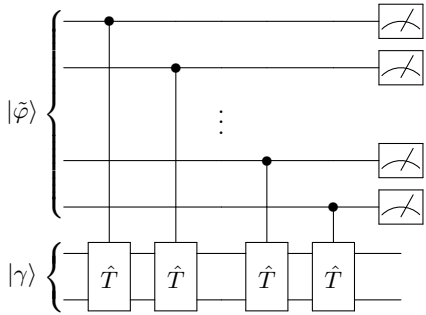

Suppose we have a controlled cyclic gate $\hat T$ working on an arbitrary number of qubits. In order to create a circuit for exponentiating such an operator we must first Taylor expand the exponential
\begin{align*}
	e^{i\theta \hat T} =& \sum_{j=0}^{\infty} \frac{1}{(nj)!}(i\theta)^{nj}\unity + \sum_{j=0}^{\infty} \frac{1}{(nj+1)!}(i\theta)^{nj+1}\hat T+\cdots\\
	&+\sum_{j=0}^{\infty} \frac{1}{((n+1)j-1)!}(i\theta)^{((n+1)j-1)}\hat T^{j-1}\\
	=& \sum_{k=0}^{n-1}\sum_{j=0}^{\infty} \frac{1}{(nj+k)!}(i\theta)^{nj+k}\hat T^k.
\end{align*}
In total this yields $n$ Taylor terms. This means that our quantum circuit would need $n-1$ ancilla qubits to perform the controls. We then apply the controlled gate $n-1$ times, each time controlled be a different ancilla qubit. The quantum circuit can be seen in \cref{fig:qcircuit}.

We must now prepare the ancilla qubits in the state
\begin{equation}
\ket{\tilde{\varphi}} = N\sum_{k=0}^{n-1}\sum_{j=0}^{\infty} \frac{1}{(nj+k)!}(i\theta)^{nj+k} \ket{\tilde{k}},
\end{equation}
where $N$ is a normalization which depends on $\theta$, and the state $\ket{\tilde{k}}$ indicates a state with $k$ excitations, i.e. we have $\ket{\tilde{0}} = \ket{00\cdots 00}$, and $\ket{\tilde{1}} = \ket{10 \cdots 00}$, $\ket{\tilde{1}} = \ket{01 \cdots 00}$, or $\ket{\tilde{1}} = \ket{00\cdots 01}$, etc. 

Let $\ket{\gamma}$ be the initial state of the target qubits. If we act with the $n-1$ controlled-$\hat T$ gates on the initial state $\ket{\tilde{\varphi}}\ket\gamma$, as in \cref{fig:qcircuit} we arrive at the state
\begin{align*}
	\ket{\tilde{\varphi}}\ket\gamma \rightarrow N\sum_{k=0}^{n-1}\sum_{j=0}^{\infty} \frac{1}{(nj+k)!}(i\theta)^{nj+k}\hat T^k \ket{\tilde{k}}\ket{\gamma}.
\end{align*}
If we measure the $n-1$ ancillae in the $\{\ket{\pm}\} = \{(\ket{0}+\ket{1})/\sqrt{2}\}$ basis, there is a probability of around $1/2^{n-1}$ that we measure $\ket{+}$ in all of the ancillae, if we require $\theta$ to be small. This means that the total state becomes
\begin{align*}
	&\ket{+\cdots +}N\sum_{k=0}^{n-1}\sum_{j=0}^{\infty} \frac{1}{(nj+k)!}(i\theta)^{nj+k}\hat T^k \ket{\gamma}\\
	&=  \ket{+\cdots +}Ne^{i\theta \hat T}\ket{\gamma},
\end{align*}
which is the desired result. If this state is not measured the experiment must be repeated until the desired result is obtained.

We note that if the gate is not cyclic our method works approximately as long as $\theta$ is small, in which case the first terms of the Taylor expansion will dominate. This means that we can chose the number of terms we want in our Taylor expansion as the number of ancillae we include in our quantum circuit.

\subsection{Example}

For an example of the Hermitian $n=2$ case see Ref. \cite{Marvian2016}. Here we consider the case $n=4$. This could, for example, be a controlled-\iswap. The exponential in this case becomes
\begin{equation}
\begin{aligned}
e^{i\theta \hat T} =& \frac{1}{2}(\cos\theta + \cosh \theta)\unity + \frac{i}{2}(\sin\theta + \sinh\theta)\hat T \\
&+ \frac{1}{2}(\cos\theta - \cosh \theta)\hat T^2 + \frac{i}{2}(\sin\theta - \sinh\theta)\hat T^3.
\end{aligned}
\end{equation}
Remember that the operator in the exponent is \emph{not} Hermitian, and thus we are \emph{not} dealing with a unitary. This means that if $\theta$ becomes large, then the hyperbolic functions will blow up. Therefore we keep $\theta$ small.
Notwithstanding, we prepare three ancillae in the state 
\begin{equation}\label{eq:phi}
\begin{aligned}
\ket{\tilde{\varphi}} =& \frac{N}{2}\big[(\cos\theta + \cosh \theta)\ket{\tilde{0}}  + i(\sin\theta + \sinh\theta)\ket{\tilde{1}} \\
&+ (\cos\theta - \cosh \theta)\ket{\tilde{2}} + i(\sin\theta - \sinh\theta)\ket{\tilde{3}}\big] \\
=& A\ket{000} + B\ket{001} + C\ket{011} + D\ket{111}.
\end{aligned}
\end{equation}
All normalization is included in $N$. Note that we could have chosen other states such as $\ket{100}$ and $\ket{101}$ in the second and third term of $\ket{\tilde{\varphi}}$ as well, as this choice can be made without loss of generality.

Now preforming the three controlled $T$-gates on the qubits we arrive at the state
\begin{align*}
	&\ket{\tilde{\varphi}}\ket{\gamma} \rightarrow \frac{N}{2}\big[(\cos\theta + \cosh \theta)\ket{\tilde{0}} + i(\sin\theta + \sinh\theta)\ket{\tilde{1}}\hat T \\
	&+ (\cos\theta - \cosh \theta)\ket{\tilde{2}}\hat T^2 + i(\sin\theta - \sinh\theta)\ket{\tilde{3}}\hat T^3\big]\ket{\gamma}.
\end{align*}
By measuring in the $\{\ket{\pm}\}$-basis there is a probability that we will measure the state $\ket{+++}$ which means that we have achieved matrix exponentiation by arriving at the state $\ket{+++}Ne^{i\theta \hat T}\ket{\gamma}$.
	 
\subsection{Measuring probabillity}\label{sec:measuringProbability}

In order to investigate the probability of measuring the correct state, we consider the state $\ket{\tilde{\varphi}}$ in \cref{eq:phi}. In the $\{\ket{\pm}\}$-basis it takes the form
\begin{equation}\label{eq:phipm}
\begin{aligned}
\ket{\tilde{\varphi}} &= \left[A + B + C + D\right] \ket{+++}\\
&+ \left[A - B - C - D\right] \ket{++-} \\
&+ \left[A + B - C - D\right] \ket{+-+} \\
&+ \left[A + B + C - D\right] \ket{-++} \\
&+ \left[A - B + C + D\right] \ket{+--} \\
&+ \left[A - B - C + D\right] \ket{-+-} \\
&+ \left[A + B - C + D\right] \ket{--+} \\
&+ \left[A - B + C - D\right] \ket{---}, \\
\end{aligned}
\end{equation}
We wish to measure a state with a coefficient $A+B+C+D$, and thus we want to measure the state $\ket{+++}$. Note that if we chose our $\ket{\tilde{k}}$ states as superpositions, such as $\ket{\tilde{1}} = a\ket{001} + b \ket{010} + c\ket{100}$, then there is no state in the $\{\ket{\pm}\}$-basis with a coefficient $A+B+C+D$, since the normalization then require the $B$ and $C$ coefficients to be normalized by the superposition coefficients $a$, $b$, and $c$, which means that we get an imbalance between the $B$ and $C$ coefficients and the $A$ and $D$ coefficients. 

\begin{figure}
	\centering
\includegraphics[width=0.9\columnwidth]{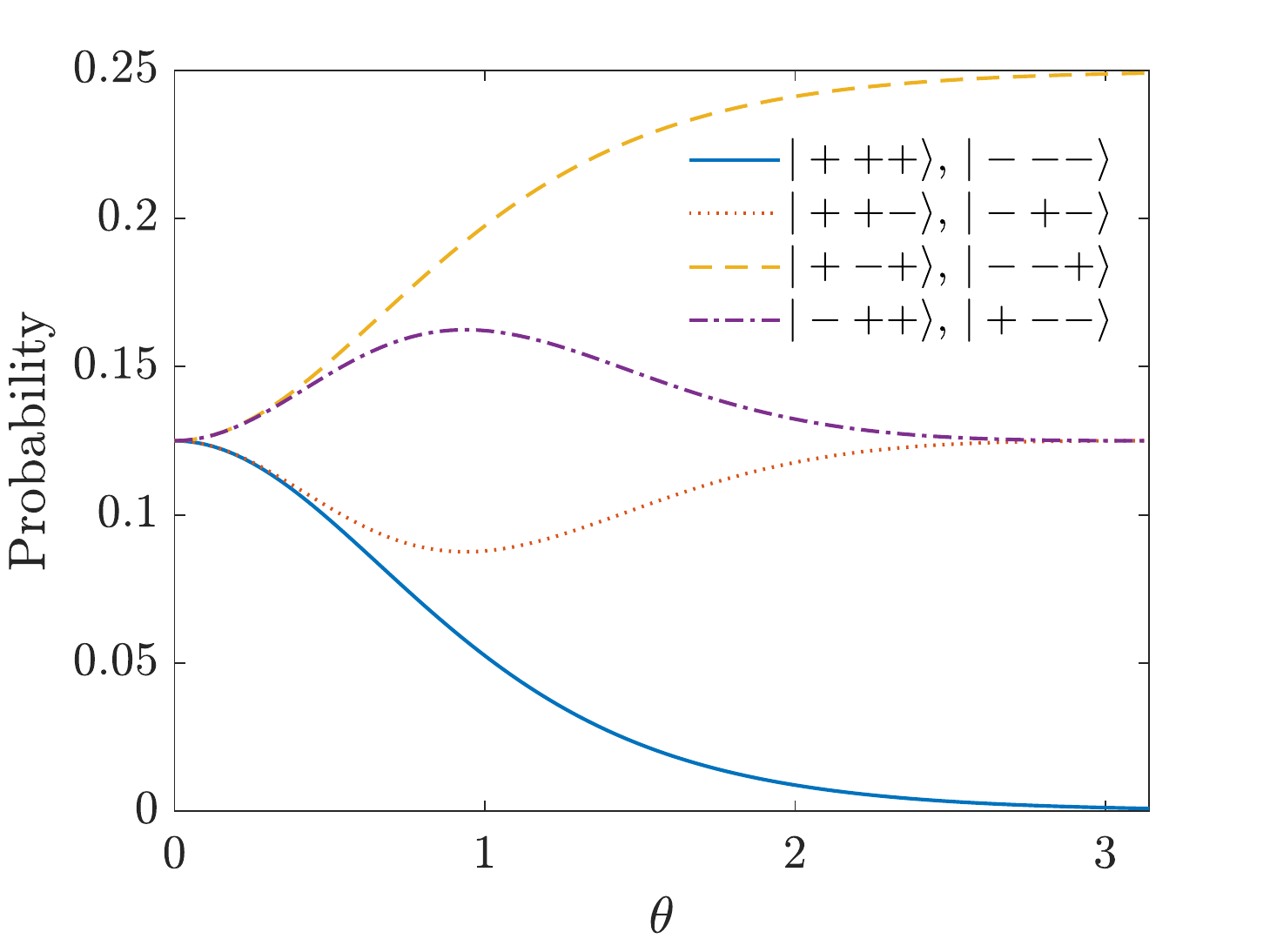}
	\caption{Probability of measuring each of the states in \cref{eq:phipm}. }
	\label{fig:prob}
\end{figure}

We plot the probabilities of measuring the eight states as a function of $\theta$ to see how they behave. The result is seen in \cref{fig:prob}.
Unfortunately we observe that the probability of measuring the state $\ket{+++}$ decreases exponentially with $\theta$. This supports our previous understanding that we should indeed keep $\theta$ small.
	 
\section{Conclusion}\label{sec:conclusion}

We have proposed a simple implementation of a controlled \iswap-gate, and shown that these exhibit a high fidelity. We have discussed an implementation of our gates using superconducting circuits and simulated the gate including possible fabrication errors and decoherence noise. Even when including these we still find a reasonably high fidelity. The general implementation presented in \cref{sec:implementation} is, however, not limited to superconducting circuits. While the difficulty of implementing our gates does increase with the number of controls, we do believe that our gates will be superior in certain types of quantum computations, especially compared to equivalent circuits built from one- and two-qubit gates, which often become quite deep. Our controlled-\iswap can easily be extended to swapping between more qubits, such that it is possible to control swapping between three, four and so on qubits. We also propose a quantum circuit for probabilistic exponentiating of non-Hermitian quantum gates, which is exact for cyclic gates and approximately exact given small parameters for all other non-Hermitian gates. These results could enhance the performance of near-term quantum computing experiments on algorithms that require multi-qubit swapping gates and exponentiating of gates.
	
\begin{acknowledgements}
	The authors thank N. J. S Loft, K. S. Christensen, T. Bækkegaard, and L. B. Kristensen for discussion on different aspects of the work.
	This work is supported by the Danish Council for Independent Research and the Carlsberg Foundation.
\end{acknowledgements}


\begin{thebibliography}{66}%
	\makeatletter
	\providecommand \@ifxundefined [1]{%
		\@ifx{#1\undefined}
	}%
	\providecommand \@ifnum [1]{%
		\ifnum #1\expandafter \@firstoftwo
		\else \expandafter \@secondoftwo
		\fi
	}%
	\providecommand \@ifx [1]{%
		\ifx #1\expandafter \@firstoftwo
		\else \expandafter \@secondoftwo
		\fi
	}%
	\providecommand \natexlab [1]{#1}%
	\providecommand \enquote  [1]{``#1''}%
	\providecommand \bibnamefont  [1]{#1}%
	\providecommand \bibfnamefont [1]{#1}%
	\providecommand \citenamefont [1]{#1}%
	\providecommand \href@noop [0]{\@secondoftwo}%
	\providecommand \href [0]{\begingroup \@sanitize@url \@href}%
	\providecommand \@href[1]{\@@startlink{#1}\@@href}%
	\providecommand \@@href[1]{\endgroup#1\@@endlink}%
	\providecommand \@sanitize@url [0]{\catcode `\\12\catcode `\$12\catcode
		`\&12\catcode `\#12\catcode `\^12\catcode `\_12\catcode `\%12\relax}%
	\providecommand \@@startlink[1]{}%
	\providecommand \@@endlink[0]{}%
	\providecommand \url  [0]{\begingroup\@sanitize@url \@url }%
	\providecommand \@url [1]{\endgroup\@href {#1}{\urlprefix }}%
	\providecommand \urlprefix  [0]{URL }%
	\providecommand \Eprint [0]{\href }%
	\providecommand \doibase [0]{http://dx.doi.org/}%
	\providecommand \selectlanguage [0]{\@gobble}%
	\providecommand \bibinfo  [0]{\@secondoftwo}%
	\providecommand \bibfield  [0]{\@secondoftwo}%
	\providecommand \translation [1]{[#1]}%
	\providecommand \BibitemOpen [0]{}%
	\providecommand \bibitemStop [0]{}%
	\providecommand \bibitemNoStop [0]{.\EOS\space}%
	\providecommand \EOS [0]{\spacefactor3000\relax}%
	\providecommand \BibitemShut  [1]{\csname bibitem#1\endcsname}%
	\let\auto@bib@innerbib\@empty
	\bibitem [{\citenamefont {Nielsen}(2002)}]{Nielsen2002}%
	\BibitemOpen
	\bibfield  {author} {\bibinfo {author} {\bibfnamefont {M.~A.}\ \bibnamefont
			{Nielsen}},\ }\href {\doibase https://doi.org/10.1016/S0375-9601(02)01272-0}
	{\bibfield  {journal} {\bibinfo  {journal} {Physics Letters A}\ }\textbf
		{\bibinfo {volume} {303}},\ \bibinfo {pages} {249 } (\bibinfo {year}
		{2002})}\BibitemShut {NoStop}%
	\bibitem [{\citenamefont {Vandersypen}\ \emph {et~al.}(2001)\citenamefont
		{Vandersypen}, \citenamefont {Steffen}, \citenamefont {Breyta}, \citenamefont
		{Yannoni}, \citenamefont {Sherwood},\ and\ \citenamefont
		{Chuang}}]{Vandersypen2001}%
	\BibitemOpen
	\bibfield  {author} {\bibinfo {author} {\bibfnamefont {L.~M.~K.}\
			\bibnamefont {Vandersypen}}, \bibinfo {author} {\bibfnamefont
			{M.}~\bibnamefont {Steffen}}, \bibinfo {author} {\bibfnamefont
			{G.}~\bibnamefont {Breyta}}, \bibinfo {author} {\bibfnamefont {C.~S.}\
			\bibnamefont {Yannoni}}, \bibinfo {author} {\bibfnamefont {M.~H.}\
			\bibnamefont {Sherwood}}, \ and\ \bibinfo {author} {\bibfnamefont {I.~L.}\
			\bibnamefont {Chuang}},\ }\href {\doibase 10.1038/414883a} {\bibfield
		{journal} {\bibinfo  {journal} {Nature}\ }\textbf {\bibinfo {volume} {414}},\
		\bibinfo {pages} {883} (\bibinfo {year} {2001})}\BibitemShut {NoStop}%
	\bibitem [{\citenamefont {Mart{\'i}n-L{\'o}pez}\ \emph
		{et~al.}(2012)\citenamefont {Mart{\'i}n-L{\'o}pez}, \citenamefont {Laing},
		\citenamefont {Lawson}, \citenamefont {Alvarez}, \citenamefont {Zhou},\ and\
		\citenamefont {O'Brien}}]{Martin-Lopez2012}%
	\BibitemOpen
	\bibfield  {author} {\bibinfo {author} {\bibfnamefont {E.}~\bibnamefont
			{Mart{\'i}n-L{\'o}pez}}, \bibinfo {author} {\bibfnamefont {A.}~\bibnamefont
			{Laing}}, \bibinfo {author} {\bibfnamefont {T.}~\bibnamefont {Lawson}},
		\bibinfo {author} {\bibfnamefont {R.}~\bibnamefont {Alvarez}}, \bibinfo
		{author} {\bibfnamefont {X.-Q.}\ \bibnamefont {Zhou}}, \ and\ \bibinfo
		{author} {\bibfnamefont {J.~L.}\ \bibnamefont {O'Brien}},\ }\href
	{https://doi.org/10.1038/nphoton.2012.259} {\bibfield  {journal} {\bibinfo
			{journal} {Nature Photonics}\ }\textbf {\bibinfo {volume} {6}},\ \bibinfo
		{pages} {773 EP } (\bibinfo {year} {2012})}\BibitemShut {NoStop}%
	\bibitem [{\citenamefont {Lanyon}\ \emph {et~al.}(2007)\citenamefont {Lanyon},
		\citenamefont {Weinhold}, \citenamefont {Langford}, \citenamefont {Barbieri},
		\citenamefont {James}, \citenamefont {Gilchrist},\ and\ \citenamefont
		{White}}]{Lanyon2007}%
	\BibitemOpen
	\bibfield  {author} {\bibinfo {author} {\bibfnamefont {B.~P.}\ \bibnamefont
			{Lanyon}}, \bibinfo {author} {\bibfnamefont {T.~J.}\ \bibnamefont
			{Weinhold}}, \bibinfo {author} {\bibfnamefont {N.~K.}\ \bibnamefont
			{Langford}}, \bibinfo {author} {\bibfnamefont {M.}~\bibnamefont {Barbieri}},
		\bibinfo {author} {\bibfnamefont {D.~F.~V.}\ \bibnamefont {James}}, \bibinfo
		{author} {\bibfnamefont {A.}~\bibnamefont {Gilchrist}}, \ and\ \bibinfo
		{author} {\bibfnamefont {A.~G.}\ \bibnamefont {White}},\ }\href {\doibase
		10.1103/PhysRevLett.99.250505} {\bibfield  {journal} {\bibinfo  {journal}
			{Phys. Rev. Lett.}\ }\textbf {\bibinfo {volume} {99}},\ \bibinfo {pages}
		{250505} (\bibinfo {year} {2007})}\BibitemShut {NoStop}%
	\bibitem [{\citenamefont {Chuang}\ and\ \citenamefont
		{Yamamoto}(1996)}]{Chuang1996}%
	\BibitemOpen
	\bibfield  {author} {\bibinfo {author} {\bibfnamefont {I.~L.}\ \bibnamefont
			{Chuang}}\ and\ \bibinfo {author} {\bibfnamefont {Y.}~\bibnamefont
			{Yamamoto}},\ }\href {\doibase 10.1103/PhysRevLett.76.4281} {\bibfield
		{journal} {\bibinfo  {journal} {Phys. Rev. Lett.}\ }\textbf {\bibinfo
			{volume} {76}},\ \bibinfo {pages} {4281} (\bibinfo {year}
		{1996})}\BibitemShut {NoStop}%
	\bibitem [{\citenamefont {Barenco}\ \emph {et~al.}(1997)\citenamefont
		{Barenco}, \citenamefont {Berthiaume}, \citenamefont {Deutsch}, \citenamefont
		{Ekert}, \citenamefont {Jozsa},\ and\ \citenamefont
		{Macchiavello}}]{Barenco1997}%
	\BibitemOpen
	\bibfield  {author} {\bibinfo {author} {\bibfnamefont {A.}~\bibnamefont
			{Barenco}}, \bibinfo {author} {\bibfnamefont {A.}~\bibnamefont {Berthiaume}},
		\bibinfo {author} {\bibfnamefont {D.}~\bibnamefont {Deutsch}}, \bibinfo
		{author} {\bibfnamefont {A.}~\bibnamefont {Ekert}}, \bibinfo {author}
		{\bibfnamefont {R.}~\bibnamefont {Jozsa}}, \ and\ \bibinfo {author}
		{\bibfnamefont {C.}~\bibnamefont {Macchiavello}},\ }\href {\doibase
		10.1137/S0097539796302452} {\bibfield  {journal} {\bibinfo  {journal} {SIAM
				Journal on Computing}\ }\textbf {\bibinfo {volume} {26}},\ \bibinfo {pages}
		{1541} (\bibinfo {year} {1997})}\BibitemShut {NoStop}%
	\bibitem [{\citenamefont {Cory}\ \emph {et~al.}(1998)\citenamefont {Cory},
		\citenamefont {Price}, \citenamefont {Maas}, \citenamefont {Knill},
		\citenamefont {Laflamme}, \citenamefont {Zurek}, \citenamefont {Havel},\ and\
		\citenamefont {Somaroo}}]{Cory1998}%
	\BibitemOpen
	\bibfield  {author} {\bibinfo {author} {\bibfnamefont {D.~G.}\ \bibnamefont
			{Cory}}, \bibinfo {author} {\bibfnamefont {M.~D.}\ \bibnamefont {Price}},
		\bibinfo {author} {\bibfnamefont {W.}~\bibnamefont {Maas}}, \bibinfo {author}
		{\bibfnamefont {E.}~\bibnamefont {Knill}}, \bibinfo {author} {\bibfnamefont
			{R.}~\bibnamefont {Laflamme}}, \bibinfo {author} {\bibfnamefont {W.~H.}\
			\bibnamefont {Zurek}}, \bibinfo {author} {\bibfnamefont {T.~F.}\ \bibnamefont
			{Havel}}, \ and\ \bibinfo {author} {\bibfnamefont {S.~S.}\ \bibnamefont
			{Somaroo}},\ }\href {\doibase 10.1103/PhysRevLett.81.2152} {\bibfield
		{journal} {\bibinfo  {journal} {Phys. Rev. Lett.}\ }\textbf {\bibinfo
			{volume} {81}},\ \bibinfo {pages} {2152} (\bibinfo {year}
		{1998})}\BibitemShut {NoStop}%
	\bibitem [{\citenamefont {Schindler}\ \emph {et~al.}(2011)\citenamefont
		{Schindler}, \citenamefont {Barreiro}, \citenamefont {Monz}, \citenamefont
		{Nebendahl}, \citenamefont {Nigg}, \citenamefont {Chwalla}, \citenamefont
		{Hennrich},\ and\ \citenamefont {Blatt}}]{Schindler2011}%
	\BibitemOpen
	\bibfield  {author} {\bibinfo {author} {\bibfnamefont {P.}~\bibnamefont
			{Schindler}}, \bibinfo {author} {\bibfnamefont {J.~T.}\ \bibnamefont
			{Barreiro}}, \bibinfo {author} {\bibfnamefont {T.}~\bibnamefont {Monz}},
		\bibinfo {author} {\bibfnamefont {V.}~\bibnamefont {Nebendahl}}, \bibinfo
		{author} {\bibfnamefont {D.}~\bibnamefont {Nigg}}, \bibinfo {author}
		{\bibfnamefont {M.}~\bibnamefont {Chwalla}}, \bibinfo {author} {\bibfnamefont
			{M.}~\bibnamefont {Hennrich}}, \ and\ \bibinfo {author} {\bibfnamefont
			{R.}~\bibnamefont {Blatt}},\ }\href {\doibase 10.1126/science.1203329}
	{\bibfield  {journal} {\bibinfo  {journal} {Science}\ }\textbf {\bibinfo
			{volume} {332}},\ \bibinfo {pages} {1059} (\bibinfo {year}
		{2011})}\BibitemShut {NoStop}%
	\bibitem [{\citenamefont {Buhrman}\ \emph {et~al.}(2001)\citenamefont
		{Buhrman}, \citenamefont {Cleve}, \citenamefont {Watrous},\ and\
		\citenamefont {de~Wolf}}]{Buhrman2001}%
	\BibitemOpen
	\bibfield  {author} {\bibinfo {author} {\bibfnamefont {H.}~\bibnamefont
			{Buhrman}}, \bibinfo {author} {\bibfnamefont {R.}~\bibnamefont {Cleve}},
		\bibinfo {author} {\bibfnamefont {J.}~\bibnamefont {Watrous}}, \ and\
		\bibinfo {author} {\bibfnamefont {R.}~\bibnamefont {de~Wolf}},\ }\href
	{\doibase 10.1103/PhysRevLett.87.167902} {\bibfield  {journal} {\bibinfo
			{journal} {Phys. Rev. Lett.}\ }\textbf {\bibinfo {volume} {87}},\ \bibinfo
		{pages} {167902} (\bibinfo {year} {2001})}\BibitemShut {NoStop}%
	\bibitem [{\citenamefont {Horn}\ \emph {et~al.}(2005)\citenamefont {Horn},
		\citenamefont {Babichev}, \citenamefont {Marzlin}, \citenamefont {Lvovsky},\
		and\ \citenamefont {Sanders}}]{Horn2005}%
	\BibitemOpen
	\bibfield  {author} {\bibinfo {author} {\bibfnamefont {R.~T.}\ \bibnamefont
			{Horn}}, \bibinfo {author} {\bibfnamefont {S.~A.}\ \bibnamefont {Babichev}},
		\bibinfo {author} {\bibfnamefont {K.-P.}\ \bibnamefont {Marzlin}}, \bibinfo
		{author} {\bibfnamefont {A.~I.}\ \bibnamefont {Lvovsky}}, \ and\ \bibinfo
		{author} {\bibfnamefont {B.~C.}\ \bibnamefont {Sanders}},\ }\href {\doibase
		10.1103/PhysRevLett.95.150502} {\bibfield  {journal} {\bibinfo  {journal}
			{Phys. Rev. Lett.}\ }\textbf {\bibinfo {volume} {95}},\ \bibinfo {pages}
		{150502} (\bibinfo {year} {2005})}\BibitemShut {NoStop}%
	\bibitem [{\citenamefont {Gottesman}\ and\ \citenamefont
		{Chuang}(2001)}]{Gottesman2001}%
	\BibitemOpen
	\bibfield  {author} {\bibinfo {author} {\bibfnamefont {D.}~\bibnamefont
			{Gottesman}}\ and\ \bibinfo {author} {\bibfnamefont {I.}~\bibnamefont
			{Chuang}},\ }\href@noop {} {\enquote {\bibinfo {title} {Quantum digital
				signatures},}\ } (\bibinfo {year} {2001}),\ \bibinfo {note}
	{arXiv:quant-ph/0105032}\BibitemShut {NoStop}%
	\bibitem [{\citenamefont {Dennis}(2001)}]{Dennis2001}%
	\BibitemOpen
	\bibfield  {author} {\bibinfo {author} {\bibfnamefont {E.}~\bibnamefont
			{Dennis}},\ }\href {\doibase 10.1103/PhysRevA.63.052314} {\bibfield
		{journal} {\bibinfo  {journal} {Phys. Rev. A}\ }\textbf {\bibinfo {volume}
			{63}},\ \bibinfo {pages} {052314} (\bibinfo {year} {2001})}\BibitemShut
	{NoStop}%
	\bibitem [{\citenamefont {Paetznick}\ and\ \citenamefont
		{Reichardt}(2013)}]{Paetznick2013}%
	\BibitemOpen
	\bibfield  {author} {\bibinfo {author} {\bibfnamefont {A.}~\bibnamefont
			{Paetznick}}\ and\ \bibinfo {author} {\bibfnamefont {B.~W.}\ \bibnamefont
			{Reichardt}},\ }\href {\doibase 10.1103/PhysRevLett.111.090505} {\bibfield
		{journal} {\bibinfo  {journal} {Phys. Rev. Lett.}\ }\textbf {\bibinfo
			{volume} {111}},\ \bibinfo {pages} {090505} (\bibinfo {year}
		{2013})}\BibitemShut {NoStop}%
	\bibitem [{\citenamefont {Ekert}\ \emph {et~al.}(2002)\citenamefont {Ekert},
		\citenamefont {Alves}, \citenamefont {Oi}, \citenamefont {Horodecki},
		\citenamefont {Horodecki},\ and\ \citenamefont {Kwek}}]{Ekert2002}%
	\BibitemOpen
	\bibfield  {author} {\bibinfo {author} {\bibfnamefont {A.~K.}\ \bibnamefont
			{Ekert}}, \bibinfo {author} {\bibfnamefont {C.~M.}\ \bibnamefont {Alves}},
		\bibinfo {author} {\bibfnamefont {D.~K.~L.}\ \bibnamefont {Oi}}, \bibinfo
		{author} {\bibfnamefont {M.}~\bibnamefont {Horodecki}}, \bibinfo {author}
		{\bibfnamefont {P.}~\bibnamefont {Horodecki}}, \ and\ \bibinfo {author}
		{\bibfnamefont {L.~C.}\ \bibnamefont {Kwek}},\ }\href {\doibase
		10.1103/PhysRevLett.88.217901} {\bibfield  {journal} {\bibinfo  {journal}
			{Phys. Rev. Lett.}\ }\textbf {\bibinfo {volume} {88}},\ \bibinfo {pages}
		{217901} (\bibinfo {year} {2002})}\BibitemShut {NoStop}%
	\bibitem [{\citenamefont {Fiur\'a\ifmmode~\check{s}\else \v{s}\fi{}ek}\ \emph
		{et~al.}(2002)\citenamefont {Fiur\'a\ifmmode~\check{s}\else \v{s}\fi{}ek},
		\citenamefont {Du\ifmmode~\check{s}\else \v{s}\fi{}ek},\ and\ \citenamefont
		{Filip}}]{Fiuraifmmode2002}%
	\BibitemOpen
	\bibfield  {author} {\bibinfo {author} {\bibfnamefont {J.}~\bibnamefont
			{Fiur\'a\ifmmode~\check{s}\else \v{s}\fi{}ek}}, \bibinfo {author}
		{\bibfnamefont {M.}~\bibnamefont {Du\ifmmode~\check{s}\else \v{s}\fi{}ek}}, \
		and\ \bibinfo {author} {\bibfnamefont {R.}~\bibnamefont {Filip}},\ }\href
	{\doibase 10.1103/PhysRevLett.89.190401} {\bibfield  {journal} {\bibinfo
			{journal} {Phys. Rev. Lett.}\ }\textbf {\bibinfo {volume} {89}},\ \bibinfo
		{pages} {190401} (\bibinfo {year} {2002})}\BibitemShut {NoStop}%
	\bibitem [{\citenamefont {Schuch}\ and\ \citenamefont
		{Siewert}(2003)}]{Schuch2003}%
	\BibitemOpen
	\bibfield  {author} {\bibinfo {author} {\bibfnamefont {N.}~\bibnamefont
			{Schuch}}\ and\ \bibinfo {author} {\bibfnamefont {J.}~\bibnamefont
			{Siewert}},\ }\href {\doibase 10.1103/PhysRevA.67.032301} {\bibfield
		{journal} {\bibinfo  {journal} {Phys. Rev. A}\ }\textbf {\bibinfo {volume}
			{67}},\ \bibinfo {pages} {032301} (\bibinfo {year} {2003})}\BibitemShut
	{NoStop}%
	\bibitem [{\citenamefont {Tanamoto}\ \emph {et~al.}(2008)\citenamefont
		{Tanamoto}, \citenamefont {Maruyama}, \citenamefont {Liu}, \citenamefont
		{Hu},\ and\ \citenamefont {Nori}}]{Tanamoto2008}%
	\BibitemOpen
	\bibfield  {author} {\bibinfo {author} {\bibfnamefont {T.}~\bibnamefont
			{Tanamoto}}, \bibinfo {author} {\bibfnamefont {K.}~\bibnamefont {Maruyama}},
		\bibinfo {author} {\bibfnamefont {Y.-x.}\ \bibnamefont {Liu}}, \bibinfo
		{author} {\bibfnamefont {X.}~\bibnamefont {Hu}}, \ and\ \bibinfo {author}
		{\bibfnamefont {F.}~\bibnamefont {Nori}},\ }\href {\doibase
		10.1103/PhysRevA.78.062313} {\bibfield  {journal} {\bibinfo  {journal} {Phys.
				Rev. A}\ }\textbf {\bibinfo {volume} {78}},\ \bibinfo {pages} {062313}
		(\bibinfo {year} {2008})}\BibitemShut {NoStop}%
	\bibitem [{\citenamefont {Tanamoto}\ \emph {et~al.}(2009)\citenamefont
		{Tanamoto}, \citenamefont {Liu}, \citenamefont {Hu},\ and\ \citenamefont
		{Nori}}]{Tanamoto2009}%
	\BibitemOpen
	\bibfield  {author} {\bibinfo {author} {\bibfnamefont {T.}~\bibnamefont
			{Tanamoto}}, \bibinfo {author} {\bibfnamefont {Y.-x.}\ \bibnamefont {Liu}},
		\bibinfo {author} {\bibfnamefont {X.}~\bibnamefont {Hu}}, \ and\ \bibinfo
		{author} {\bibfnamefont {F.}~\bibnamefont {Nori}},\ }\href {\doibase
		10.1103/PhysRevLett.102.100501} {\bibfield  {journal} {\bibinfo  {journal}
			{Phys. Rev. Lett.}\ }\textbf {\bibinfo {volume} {102}},\ \bibinfo {pages}
		{100501} (\bibinfo {year} {2009})}\BibitemShut {NoStop}%
	\bibitem [{\citenamefont {Zagoskin}\ \emph {et~al.}(2006)\citenamefont
		{Zagoskin}, \citenamefont {Ashhab}, \citenamefont {Johansson},\ and\
		\citenamefont {Nori}}]{Zagoskin2006}%
	\BibitemOpen
	\bibfield  {author} {\bibinfo {author} {\bibfnamefont {A.~M.}\ \bibnamefont
			{Zagoskin}}, \bibinfo {author} {\bibfnamefont {S.}~\bibnamefont {Ashhab}},
		\bibinfo {author} {\bibfnamefont {J.~R.}\ \bibnamefont {Johansson}}, \ and\
		\bibinfo {author} {\bibfnamefont {F.}~\bibnamefont {Nori}},\ }\href {\doibase
		10.1103/PhysRevLett.97.077001} {\bibfield  {journal} {\bibinfo  {journal}
			{Phys. Rev. Lett.}\ }\textbf {\bibinfo {volume} {97}},\ \bibinfo {pages}
		{077001} (\bibinfo {year} {2006})}\BibitemShut {NoStop}%
	\bibitem [{\citenamefont {Imamo\={g}lu}\ \emph {et~al.}(1999)\citenamefont
		{Imamo\={g}lu}, \citenamefont {Awschalom}, \citenamefont {Burkard},
		\citenamefont {DiVincenzo}, \citenamefont {Loss}, \citenamefont {Sherwin},\
		and\ \citenamefont {Small}}]{Imamoglu1999}%
	\BibitemOpen
	\bibfield  {author} {\bibinfo {author} {\bibfnamefont {A.}~\bibnamefont
			{Imamo\={g}lu}}, \bibinfo {author} {\bibfnamefont {D.~D.}\ \bibnamefont
			{Awschalom}}, \bibinfo {author} {\bibfnamefont {G.}~\bibnamefont {Burkard}},
		\bibinfo {author} {\bibfnamefont {D.~P.}\ \bibnamefont {DiVincenzo}},
		\bibinfo {author} {\bibfnamefont {D.}~\bibnamefont {Loss}}, \bibinfo {author}
		{\bibfnamefont {M.}~\bibnamefont {Sherwin}}, \ and\ \bibinfo {author}
		{\bibfnamefont {A.}~\bibnamefont {Small}},\ }\href {\doibase
		10.1103/PhysRevLett.83.4204} {\bibfield  {journal} {\bibinfo  {journal}
			{Phys. Rev. Lett.}\ }\textbf {\bibinfo {volume} {83}},\ \bibinfo {pages}
		{4204} (\bibinfo {year} {1999})}\BibitemShut {NoStop}%
	\bibitem [{\citenamefont {Benito}\ \emph {et~al.}(2019)\citenamefont {Benito},
		\citenamefont {Petta},\ and\ \citenamefont {Burkard}}]{Benito2019}%
	\BibitemOpen
	\bibfield  {author} {\bibinfo {author} {\bibfnamefont {M.}~\bibnamefont
			{Benito}}, \bibinfo {author} {\bibfnamefont {J.~R.}\ \bibnamefont {Petta}}, \
		and\ \bibinfo {author} {\bibfnamefont {G.}~\bibnamefont {Burkard}},\ }\href
	{\doibase 10.1103/PhysRevB.100.081412} {\bibfield  {journal} {\bibinfo
			{journal} {Phys. Rev. B}\ }\textbf {\bibinfo {volume} {100}},\ \bibinfo
		{pages} {081412(R)} (\bibinfo {year} {2019})}\BibitemShut {NoStop}%
	\bibitem [{\citenamefont {Blais}\ \emph {et~al.}(2004)\citenamefont {Blais},
		\citenamefont {Huang}, \citenamefont {Wallraff}, \citenamefont {Girvin},\
		and\ \citenamefont {Schoelkopf}}]{Blais2004}%
	\BibitemOpen
	\bibfield  {author} {\bibinfo {author} {\bibfnamefont {A.}~\bibnamefont
			{Blais}}, \bibinfo {author} {\bibfnamefont {R.-S.}\ \bibnamefont {Huang}},
		\bibinfo {author} {\bibfnamefont {A.}~\bibnamefont {Wallraff}}, \bibinfo
		{author} {\bibfnamefont {S.~M.}\ \bibnamefont {Girvin}}, \ and\ \bibinfo
		{author} {\bibfnamefont {R.~J.}\ \bibnamefont {Schoelkopf}},\ }\href
	{\doibase 10.1103/PhysRevA.69.062320} {\bibfield  {journal} {\bibinfo
			{journal} {Phys. Rev. A}\ }\textbf {\bibinfo {volume} {69}},\ \bibinfo
		{pages} {062320} (\bibinfo {year} {2004})}\BibitemShut {NoStop}%
	\bibitem [{\citenamefont {Wang}\ \emph {et~al.}(2010)\citenamefont {Wang},
		\citenamefont {Shao}, \citenamefont {Zhao}, \citenamefont {Zhang},\ and\
		\citenamefont {Yeon}}]{Wang2010}%
	\BibitemOpen
	\bibfield  {author} {\bibinfo {author} {\bibfnamefont {H.-F.}\ \bibnamefont
			{Wang}}, \bibinfo {author} {\bibfnamefont {X.-Q.}\ \bibnamefont {Shao}},
		\bibinfo {author} {\bibfnamefont {Y.-F.}\ \bibnamefont {Zhao}}, \bibinfo
		{author} {\bibfnamefont {S.}~\bibnamefont {Zhang}}, \ and\ \bibinfo {author}
		{\bibfnamefont {K.-H.}\ \bibnamefont {Yeon}},\ }\href {\doibase
		10.1364/JOSAB.27.000027} {\bibfield  {journal} {\bibinfo  {journal} {J. Opt.
				Soc. Am. B}\ }\textbf {\bibinfo {volume} {27}},\ \bibinfo {pages} {27}
		(\bibinfo {year} {2010})}\BibitemShut {NoStop}%
	\bibitem [{\citenamefont {Bartkowiak}\ and\ \citenamefont
		{Miranowicz}(2010)}]{Bartkowiak2010}%
	\BibitemOpen
	\bibfield  {author} {\bibinfo {author} {\bibfnamefont {M.}~\bibnamefont
			{Bartkowiak}}\ and\ \bibinfo {author} {\bibfnamefont {A.}~\bibnamefont
			{Miranowicz}},\ }\href {\doibase 10.1364/JOSAB.27.002369} {\bibfield
		{journal} {\bibinfo  {journal} {J. Opt. Soc. Am. B}\ }\textbf {\bibinfo
			{volume} {27}},\ \bibinfo {pages} {2369} (\bibinfo {year}
		{2010})}\BibitemShut {NoStop}%
	\bibitem [{\citenamefont {Godfrin}\ \emph {et~al.}(2018)\citenamefont
		{Godfrin}, \citenamefont {Ballou}, \citenamefont {Bonet}, \citenamefont
		{Ruben}, \citenamefont {Klyatskaya}, \citenamefont {Wernsdorfer},\ and\
		\citenamefont {Balestro}}]{Godfrin2018}%
	\BibitemOpen
	\bibfield  {author} {\bibinfo {author} {\bibfnamefont {C.}~\bibnamefont
			{Godfrin}}, \bibinfo {author} {\bibfnamefont {R.}~\bibnamefont {Ballou}},
		\bibinfo {author} {\bibfnamefont {E.}~\bibnamefont {Bonet}}, \bibinfo
		{author} {\bibfnamefont {M.}~\bibnamefont {Ruben}}, \bibinfo {author}
		{\bibfnamefont {S.}~\bibnamefont {Klyatskaya}}, \bibinfo {author}
		{\bibfnamefont {W.}~\bibnamefont {Wernsdorfer}}, \ and\ \bibinfo {author}
		{\bibfnamefont {F.}~\bibnamefont {Balestro}},\ }\href {\doibase
		10.1038/s41534-018-0101-3} {\bibfield  {journal} {\bibinfo  {journal} {npj
				Quantum Information}\ }\textbf {\bibinfo {volume} {4}},\ \bibinfo {pages}
		{53} (\bibinfo {year} {2018})}\BibitemShut {NoStop}%
	\bibitem [{\citenamefont {McKay}\ \emph {et~al.}(2016)\citenamefont {McKay},
		\citenamefont {Filipp}, \citenamefont {Mezzacapo}, \citenamefont {Magesan},
		\citenamefont {Chow},\ and\ \citenamefont {Gambetta}}]{McKay2016}%
	\BibitemOpen
	\bibfield  {author} {\bibinfo {author} {\bibfnamefont {D.~C.}\ \bibnamefont
			{McKay}}, \bibinfo {author} {\bibfnamefont {S.}~\bibnamefont {Filipp}},
		\bibinfo {author} {\bibfnamefont {A.}~\bibnamefont {Mezzacapo}}, \bibinfo
		{author} {\bibfnamefont {E.}~\bibnamefont {Magesan}}, \bibinfo {author}
		{\bibfnamefont {J.~M.}\ \bibnamefont {Chow}}, \ and\ \bibinfo {author}
		{\bibfnamefont {J.~M.}\ \bibnamefont {Gambetta}},\ }\href {\doibase
		10.1103/PhysRevApplied.6.064007} {\bibfield  {journal} {\bibinfo  {journal}
			{Phys. Rev. Applied}\ }\textbf {\bibinfo {volume} {6}},\ \bibinfo {pages}
		{064007} (\bibinfo {year} {2016})}\BibitemShut {NoStop}%
	\bibitem [{\citenamefont {Dewes}\ \emph {et~al.}(2012)\citenamefont {Dewes},
		\citenamefont {Ong}, \citenamefont {Schmitt}, \citenamefont {Lauro},
		\citenamefont {Boulant}, \citenamefont {Bertet}, \citenamefont {Vion},\ and\
		\citenamefont {Esteve}}]{Dewes2012}%
	\BibitemOpen
	\bibfield  {author} {\bibinfo {author} {\bibfnamefont {A.}~\bibnamefont
			{Dewes}}, \bibinfo {author} {\bibfnamefont {F.~R.}\ \bibnamefont {Ong}},
		\bibinfo {author} {\bibfnamefont {V.}~\bibnamefont {Schmitt}}, \bibinfo
		{author} {\bibfnamefont {R.}~\bibnamefont {Lauro}}, \bibinfo {author}
		{\bibfnamefont {N.}~\bibnamefont {Boulant}}, \bibinfo {author} {\bibfnamefont
			{P.}~\bibnamefont {Bertet}}, \bibinfo {author} {\bibfnamefont
			{D.}~\bibnamefont {Vion}}, \ and\ \bibinfo {author} {\bibfnamefont
			{D.}~\bibnamefont {Esteve}},\ }\href {\doibase
		10.1103/PhysRevLett.108.057002} {\bibfield  {journal} {\bibinfo  {journal}
			{Phys. Rev. Lett.}\ }\textbf {\bibinfo {volume} {108}},\ \bibinfo {pages}
		{057002} (\bibinfo {year} {2012})}\BibitemShut {NoStop}%
	\bibitem [{\citenamefont {Salath\'e}\ \emph {et~al.}(2015)\citenamefont
		{Salath\'e}, \citenamefont {Mondal}, \citenamefont {Oppliger}, \citenamefont
		{Heinsoo}, \citenamefont {Kurpiers}, \citenamefont
		{Poto\ifmmode~\check{c}\else \v{c}\fi{}nik}, \citenamefont {Mezzacapo},
		\citenamefont {Las~Heras}, \citenamefont {Lamata}, \citenamefont {Solano},
		\citenamefont {Filipp},\ and\ \citenamefont {Wallraff}}]{Salathe2015}%
	\BibitemOpen
	\bibfield  {author} {\bibinfo {author} {\bibfnamefont {Y.}~\bibnamefont
			{Salath\'e}}, \bibinfo {author} {\bibfnamefont {M.}~\bibnamefont {Mondal}},
		\bibinfo {author} {\bibfnamefont {M.}~\bibnamefont {Oppliger}}, \bibinfo
		{author} {\bibfnamefont {J.}~\bibnamefont {Heinsoo}}, \bibinfo {author}
		{\bibfnamefont {P.}~\bibnamefont {Kurpiers}}, \bibinfo {author}
		{\bibfnamefont {A.}~\bibnamefont {Poto\ifmmode~\check{c}\else
				\v{c}\fi{}nik}}, \bibinfo {author} {\bibfnamefont {A.}~\bibnamefont
			{Mezzacapo}}, \bibinfo {author} {\bibfnamefont {U.}~\bibnamefont
			{Las~Heras}}, \bibinfo {author} {\bibfnamefont {L.}~\bibnamefont {Lamata}},
		\bibinfo {author} {\bibfnamefont {E.}~\bibnamefont {Solano}}, \bibinfo
		{author} {\bibfnamefont {S.}~\bibnamefont {Filipp}}, \ and\ \bibinfo {author}
		{\bibfnamefont {A.}~\bibnamefont {Wallraff}},\ }\href {\doibase
		10.1103/PhysRevX.5.021027} {\bibfield  {journal} {\bibinfo  {journal} {Phys.
				Rev. X}\ }\textbf {\bibinfo {volume} {5}},\ \bibinfo {pages} {021027}
		(\bibinfo {year} {2015})}\BibitemShut {NoStop}%
	\bibitem [{\citenamefont {Milburn}(1989)}]{Milburn1989}%
	\BibitemOpen
	\bibfield  {author} {\bibinfo {author} {\bibfnamefont {G.~J.}\ \bibnamefont
			{Milburn}},\ }\href {\doibase 10.1103/PhysRevLett.62.2124} {\bibfield
		{journal} {\bibinfo  {journal} {Phys. Rev. Lett.}\ }\textbf {\bibinfo
			{volume} {62}},\ \bibinfo {pages} {2124} (\bibinfo {year}
		{1989})}\BibitemShut {NoStop}%
	\bibitem [{\citenamefont {Chau}\ and\ \citenamefont
		{Wilczek}(1995)}]{Chau1995}%
	\BibitemOpen
	\bibfield  {author} {\bibinfo {author} {\bibfnamefont {H.~F.}\ \bibnamefont
			{Chau}}\ and\ \bibinfo {author} {\bibfnamefont {F.}~\bibnamefont {Wilczek}},\
	}\href {\doibase 10.1103/PhysRevLett.75.748} {\bibfield  {journal} {\bibinfo
			{journal} {Phys. Rev. Lett.}\ }\textbf {\bibinfo {volume} {75}},\ \bibinfo
		{pages} {748} (\bibinfo {year} {1995})}\BibitemShut {NoStop}%
	\bibitem [{\citenamefont {Fiur\'a\ifmmode~\check{s}\else
			\v{s}\fi{}ek}(2006)}]{Fiuraifmmode2006}%
	\BibitemOpen
	\bibfield  {author} {\bibinfo {author} {\bibfnamefont {J.}~\bibnamefont
			{Fiur\'a\ifmmode~\check{s}\else \v{s}\fi{}ek}},\ }\href {\doibase
		10.1103/PhysRevA.73.062313} {\bibfield  {journal} {\bibinfo  {journal} {Phys.
				Rev. A}\ }\textbf {\bibinfo {volume} {73}},\ \bibinfo {pages} {062313}
		(\bibinfo {year} {2006})}\BibitemShut {NoStop}%
	\bibitem [{\citenamefont {Fiur\'a\ifmmode~\check{s}\else
			\v{s}\fi{}ek}(2008)}]{Fiuraifmmode2008}%
	\BibitemOpen
	\bibfield  {author} {\bibinfo {author} {\bibfnamefont {J.}~\bibnamefont
			{Fiur\'a\ifmmode~\check{s}\else \v{s}\fi{}ek}},\ }\href {\doibase
		10.1103/PhysRevA.78.032317} {\bibfield  {journal} {\bibinfo  {journal} {Phys.
				Rev. A}\ }\textbf {\bibinfo {volume} {78}},\ \bibinfo {pages} {032317}
		(\bibinfo {year} {2008})}\BibitemShut {NoStop}%
	\bibitem [{\citenamefont {Gong}\ \emph {et~al.}(2008)\citenamefont {Gong},
		\citenamefont {Guo},\ and\ \citenamefont {Ralph}}]{Gong2008}%
	\BibitemOpen
	\bibfield  {author} {\bibinfo {author} {\bibfnamefont {Y.-X.}\ \bibnamefont
			{Gong}}, \bibinfo {author} {\bibfnamefont {G.-C.}\ \bibnamefont {Guo}}, \
		and\ \bibinfo {author} {\bibfnamefont {T.~C.}\ \bibnamefont {Ralph}},\ }\href
	{\doibase 10.1103/PhysRevA.78.012305} {\bibfield  {journal} {\bibinfo
			{journal} {Phys. Rev. A}\ }\textbf {\bibinfo {volume} {78}},\ \bibinfo
		{pages} {012305} (\bibinfo {year} {2008})}\BibitemShut {NoStop}%
	\bibitem [{\citenamefont {Patel}\ \emph {et~al.}(2016)\citenamefont {Patel},
		\citenamefont {Ho}, \citenamefont {Ferreyrol}, \citenamefont {Ralph},\ and\
		\citenamefont {Pryde}}]{Patel2016}%
	\BibitemOpen
	\bibfield  {author} {\bibinfo {author} {\bibfnamefont {R.~B.}\ \bibnamefont
			{Patel}}, \bibinfo {author} {\bibfnamefont {J.}~\bibnamefont {Ho}}, \bibinfo
		{author} {\bibfnamefont {F.}~\bibnamefont {Ferreyrol}}, \bibinfo {author}
		{\bibfnamefont {T.~C.}\ \bibnamefont {Ralph}}, \ and\ \bibinfo {author}
		{\bibfnamefont {G.~J.}\ \bibnamefont {Pryde}},\ }\href {\doibase
		10.1126/sciadv.1501531} {\bibfield  {journal} {\bibinfo  {journal} {Science
				Advances}\ }\textbf {\bibinfo {volume} {2}} (\bibinfo {year} {2016}),\
		10.1126/sciadv.1501531}\BibitemShut {NoStop}%
	\bibitem [{\citenamefont {Ono}\ \emph {et~al.}(2017)\citenamefont {Ono},
		\citenamefont {Okamoto}, \citenamefont {Tanida}, \citenamefont {Hofmann},\
		and\ \citenamefont {Takeuchi}}]{Ono2017}%
	\BibitemOpen
	\bibfield  {author} {\bibinfo {author} {\bibfnamefont {T.}~\bibnamefont
			{Ono}}, \bibinfo {author} {\bibfnamefont {R.}~\bibnamefont {Okamoto}},
		\bibinfo {author} {\bibfnamefont {M.}~\bibnamefont {Tanida}}, \bibinfo
		{author} {\bibfnamefont {H.~F.}\ \bibnamefont {Hofmann}}, \ and\ \bibinfo
		{author} {\bibfnamefont {S.}~\bibnamefont {Takeuchi}},\ }\href
	{https://doi.org/10.1038/srep45353} {\bibfield  {journal} {\bibinfo
			{journal} {Scientific Reports}\ }\textbf {\bibinfo {volume} {7}},\ \bibinfo
		{pages} {45353 EP } (\bibinfo {year} {2017})},\ \bibinfo {note}
	{article}\BibitemShut {NoStop}%
	\bibitem [{\citenamefont {Smolin}\ and\ \citenamefont
		{DiVincenzo}(1996)}]{Smolin1996}%
	\BibitemOpen
	\bibfield  {author} {\bibinfo {author} {\bibfnamefont {J.~A.}\ \bibnamefont
			{Smolin}}\ and\ \bibinfo {author} {\bibfnamefont {D.~P.}\ \bibnamefont
			{DiVincenzo}},\ }\href {\doibase 10.1103/PhysRevA.53.2855} {\bibfield
		{journal} {\bibinfo  {journal} {Phys. Rev. A}\ }\textbf {\bibinfo {volume}
			{53}},\ \bibinfo {pages} {2855} (\bibinfo {year} {1996})}\BibitemShut
	{NoStop}%
	\bibitem [{\citenamefont {B{\ae}kkegaard}\ \emph {et~al.}(2019)\citenamefont
		{B{\ae}kkegaard}, \citenamefont {Kristensen}, \citenamefont {Loft},
		\citenamefont {Andersen}, \citenamefont {Petrosyan},\ and\ \citenamefont
		{Zinner}}]{Baekkegaard2019}%
	\BibitemOpen
	\bibfield  {author} {\bibinfo {author} {\bibfnamefont {T.}~\bibnamefont
			{B{\ae}kkegaard}}, \bibinfo {author} {\bibfnamefont {L.~B.}\ \bibnamefont
			{Kristensen}}, \bibinfo {author} {\bibfnamefont {N.~J.~S.}\ \bibnamefont
			{Loft}}, \bibinfo {author} {\bibfnamefont {C.~K.}\ \bibnamefont {Andersen}},
		\bibinfo {author} {\bibfnamefont {D.}~\bibnamefont {Petrosyan}}, \ and\
		\bibinfo {author} {\bibfnamefont {N.~T.}\ \bibnamefont {Zinner}},\ }\href
	{\doibase 10.1038/s41598-019-49657-1} {\bibfield  {journal} {\bibinfo
			{journal} {Scientific Reports}\ }\textbf {\bibinfo {volume} {9}},\ \bibinfo
		{pages} {13389} (\bibinfo {year} {2019})}\BibitemShut {NoStop}%
	\bibitem [{\citenamefont {Poletto}\ \emph {et~al.}(2012)\citenamefont
		{Poletto}, \citenamefont {Gambetta}, \citenamefont {Merkel}, \citenamefont
		{Smolin}, \citenamefont {Chow}, \citenamefont {C\'orcoles}, \citenamefont
		{Keefe}, \citenamefont {Rothwell}, \citenamefont {Rozen}, \citenamefont
		{Abraham}, \citenamefont {Rigetti},\ and\ \citenamefont
		{Steffen}}]{Poletto2012}%
	\BibitemOpen
	\bibfield  {author} {\bibinfo {author} {\bibfnamefont {S.}~\bibnamefont
			{Poletto}}, \bibinfo {author} {\bibfnamefont {J.~M.}\ \bibnamefont
			{Gambetta}}, \bibinfo {author} {\bibfnamefont {S.~T.}\ \bibnamefont
			{Merkel}}, \bibinfo {author} {\bibfnamefont {J.~A.}\ \bibnamefont {Smolin}},
		\bibinfo {author} {\bibfnamefont {J.~M.}\ \bibnamefont {Chow}}, \bibinfo
		{author} {\bibfnamefont {A.~D.}\ \bibnamefont {C\'orcoles}}, \bibinfo
		{author} {\bibfnamefont {G.~A.}\ \bibnamefont {Keefe}}, \bibinfo {author}
		{\bibfnamefont {M.~B.}\ \bibnamefont {Rothwell}}, \bibinfo {author}
		{\bibfnamefont {J.~R.}\ \bibnamefont {Rozen}}, \bibinfo {author}
		{\bibfnamefont {D.~W.}\ \bibnamefont {Abraham}}, \bibinfo {author}
		{\bibfnamefont {C.}~\bibnamefont {Rigetti}}, \ and\ \bibinfo {author}
		{\bibfnamefont {M.}~\bibnamefont {Steffen}},\ }\href {\doibase
		10.1103/PhysRevLett.109.240505} {\bibfield  {journal} {\bibinfo  {journal}
			{Phys. Rev. Lett.}\ }\textbf {\bibinfo {volume} {109}},\ \bibinfo {pages}
		{240505} (\bibinfo {year} {2012})}\BibitemShut {NoStop}%
	\bibitem [{\citenamefont {Rasmussen}\ \emph {et~al.}(2019)\citenamefont
		{Rasmussen}, \citenamefont {Christensen},\ and\ \citenamefont
		{Zinner}}]{Rasmussen2019}%
	\BibitemOpen
	\bibfield  {author} {\bibinfo {author} {\bibfnamefont {S.~E.}\ \bibnamefont
			{Rasmussen}}, \bibinfo {author} {\bibfnamefont {K.~S.}\ \bibnamefont
			{Christensen}}, \ and\ \bibinfo {author} {\bibfnamefont {N.~T.}\ \bibnamefont
			{Zinner}},\ }\href {\doibase 10.1103/PhysRevB.99.134508} {\bibfield
		{journal} {\bibinfo  {journal} {Phys. Rev. B}\ }\textbf {\bibinfo {volume}
			{99}},\ \bibinfo {pages} {134508} (\bibinfo {year} {2019})}\BibitemShut
	{NoStop}%
	\bibitem [{\citenamefont {Loft}\ \emph {et~al.}(2020)\citenamefont {Loft},
		\citenamefont {Kjaergaard}, \citenamefont {Kristensen}, \citenamefont
		{Andersen}, \citenamefont {Larsen}, \citenamefont {Gustavsson}, \citenamefont
		{Oliver},\ and\ \citenamefont {Zinner}}]{Loft2020}%
	\BibitemOpen
	\bibfield  {author} {\bibinfo {author} {\bibfnamefont {N.~J.~S.}\
			\bibnamefont {Loft}}, \bibinfo {author} {\bibfnamefont {M.}~\bibnamefont
			{Kjaergaard}}, \bibinfo {author} {\bibfnamefont {L.~B.}\ \bibnamefont
			{Kristensen}}, \bibinfo {author} {\bibfnamefont {C.~K.}\ \bibnamefont
			{Andersen}}, \bibinfo {author} {\bibfnamefont {T.~W.}\ \bibnamefont
			{Larsen}}, \bibinfo {author} {\bibfnamefont {S.}~\bibnamefont {Gustavsson}},
		\bibinfo {author} {\bibfnamefont {W.~D.}\ \bibnamefont {Oliver}}, \ and\
		\bibinfo {author} {\bibfnamefont {N.~T.}\ \bibnamefont {Zinner}},\ }\href
	{\doibase 10.1038/s41534-020-0275-3} {\bibfield  {journal} {\bibinfo
			{journal} {npj Quantum Information}\ }\textbf {\bibinfo {volume} {6}},\
		\bibinfo {pages} {47} (\bibinfo {year} {2020})}\BibitemShut {NoStop}%
	\bibitem [{\citenamefont {Gao}\ \emph {et~al.}(2018)\citenamefont {Gao},
		\citenamefont {Lester}, \citenamefont {Zhang}, \citenamefont {Wang},
		\citenamefont {Rosenblum}, \citenamefont {Frunzio}, \citenamefont {Jiang},
		\citenamefont {Girvin},\ and\ \citenamefont {Schoelkopf}}]{Gao2018}%
	\BibitemOpen
	\bibfield  {author} {\bibinfo {author} {\bibfnamefont {Y.~Y.}\ \bibnamefont
			{Gao}}, \bibinfo {author} {\bibfnamefont {B.~J.}\ \bibnamefont {Lester}},
		\bibinfo {author} {\bibfnamefont {Y.}~\bibnamefont {Zhang}}, \bibinfo
		{author} {\bibfnamefont {C.}~\bibnamefont {Wang}}, \bibinfo {author}
		{\bibfnamefont {S.}~\bibnamefont {Rosenblum}}, \bibinfo {author}
		{\bibfnamefont {L.}~\bibnamefont {Frunzio}}, \bibinfo {author} {\bibfnamefont
			{L.}~\bibnamefont {Jiang}}, \bibinfo {author} {\bibfnamefont {S.~M.}\
			\bibnamefont {Girvin}}, \ and\ \bibinfo {author} {\bibfnamefont {R.~J.}\
			\bibnamefont {Schoelkopf}},\ }\href {\doibase 10.1103/PhysRevX.8.021073}
	{\bibfield  {journal} {\bibinfo  {journal} {Phys. Rev. X}\ }\textbf {\bibinfo
			{volume} {8}},\ \bibinfo {pages} {021073} (\bibinfo {year}
		{2018})}\BibitemShut {NoStop}%
	\bibitem [{\citenamefont {Gao}\ \emph {et~al.}(2019)\citenamefont {Gao},
		\citenamefont {Lester}, \citenamefont {Chou}, \citenamefont {Frunzio},
		\citenamefont {Devoret}, \citenamefont {Jiang}, \citenamefont {Girvin},\ and\
		\citenamefont {Schoelkopf}}]{Gao2019}%
	\BibitemOpen
	\bibfield  {author} {\bibinfo {author} {\bibfnamefont {Y.~Y.}\ \bibnamefont
			{Gao}}, \bibinfo {author} {\bibfnamefont {B.~J.}\ \bibnamefont {Lester}},
		\bibinfo {author} {\bibfnamefont {K.~S.}\ \bibnamefont {Chou}}, \bibinfo
		{author} {\bibfnamefont {L.}~\bibnamefont {Frunzio}}, \bibinfo {author}
		{\bibfnamefont {M.~H.}\ \bibnamefont {Devoret}}, \bibinfo {author}
		{\bibfnamefont {L.}~\bibnamefont {Jiang}}, \bibinfo {author} {\bibfnamefont
			{S.~M.}\ \bibnamefont {Girvin}}, \ and\ \bibinfo {author} {\bibfnamefont
			{R.~J.}\ \bibnamefont {Schoelkopf}},\ }\href {\doibase
		10.1038/s41586-019-0970-4} {\bibfield  {journal} {\bibinfo  {journal}
			{Nature}\ }\textbf {\bibinfo {volume} {566}},\ \bibinfo {pages} {509}
		(\bibinfo {year} {2019})}\BibitemShut {NoStop}%
	\bibitem [{\citenamefont {Rasmussen}\ \emph {et~al.}(2020)\citenamefont
		{Rasmussen}, \citenamefont {Groenland}, \citenamefont {Gerritsma},
		\citenamefont {Schoutens},\ and\ \citenamefont {Zinner}}]{Rasmussen2020}%
	\BibitemOpen
	\bibfield  {author} {\bibinfo {author} {\bibfnamefont {S.~E.}\ \bibnamefont
			{Rasmussen}}, \bibinfo {author} {\bibfnamefont {K.}~\bibnamefont
			{Groenland}}, \bibinfo {author} {\bibfnamefont {R.}~\bibnamefont
			{Gerritsma}}, \bibinfo {author} {\bibfnamefont {K.}~\bibnamefont
			{Schoutens}}, \ and\ \bibinfo {author} {\bibfnamefont {N.~T.}\ \bibnamefont
			{Zinner}},\ }\href {\doibase 10.1103/PhysRevA.101.022308} {\bibfield
		{journal} {\bibinfo  {journal} {Phys. Rev. A}\ }\textbf {\bibinfo {volume}
			{101}},\ \bibinfo {pages} {022308} (\bibinfo {year} {2020})}\BibitemShut
	{NoStop}%
	\bibitem [{\citenamefont {Christensen}\ \emph {et~al.}(2020)\citenamefont
		{Christensen}, \citenamefont {Rasmussen}, \citenamefont {Petrosyan},\ and\
		\citenamefont {Zinner}}]{Christensen2020}%
	\BibitemOpen
	\bibfield  {author} {\bibinfo {author} {\bibfnamefont {K.~S.}\ \bibnamefont
			{Christensen}}, \bibinfo {author} {\bibfnamefont {S.~E.}\ \bibnamefont
			{Rasmussen}}, \bibinfo {author} {\bibfnamefont {D.}~\bibnamefont
			{Petrosyan}}, \ and\ \bibinfo {author} {\bibfnamefont {N.~T.}\ \bibnamefont
			{Zinner}},\ }\href {\doibase 10.1103/PhysRevResearch.2.013004} {\bibfield
		{journal} {\bibinfo  {journal} {Phys. Rev. Research}\ }\textbf {\bibinfo
			{volume} {2}},\ \bibinfo {pages} {013004} (\bibinfo {year}
		{2020})}\BibitemShut {NoStop}%
	\bibitem [{\citenamefont {Braunstein}\ and\ \citenamefont {van
			Loock}(2005)}]{Braunstein2005}%
	\BibitemOpen
	\bibfield  {author} {\bibinfo {author} {\bibfnamefont {S.~L.}\ \bibnamefont
			{Braunstein}}\ and\ \bibinfo {author} {\bibfnamefont {P.}~\bibnamefont {van
				Loock}},\ }\href {\doibase 10.1103/RevModPhys.77.513} {\bibfield  {journal}
		{\bibinfo  {journal} {Rev. Mod. Phys.}\ }\textbf {\bibinfo {volume} {77}},\
		\bibinfo {pages} {513} (\bibinfo {year} {2005})}\BibitemShut {NoStop}%
	\bibitem [{\citenamefont {Lau}\ and\ \citenamefont {Plenio}(2016)}]{Lau2016}%
	\BibitemOpen
	\bibfield  {author} {\bibinfo {author} {\bibfnamefont {H.-K.}\ \bibnamefont
			{Lau}}\ and\ \bibinfo {author} {\bibfnamefont {M.~B.}\ \bibnamefont
			{Plenio}},\ }\href {\doibase 10.1103/PhysRevLett.117.100501} {\bibfield
		{journal} {\bibinfo  {journal} {Phys. Rev. Lett.}\ }\textbf {\bibinfo
			{volume} {117}},\ \bibinfo {pages} {100501} (\bibinfo {year}
		{2016})}\BibitemShut {NoStop}%
	\bibitem [{\citenamefont {Lau}\ \emph {et~al.}(2017)\citenamefont {Lau},
		\citenamefont {Pooser}, \citenamefont {Siopsis},\ and\ \citenamefont
		{Weedbrook}}]{Lau2017}%
	\BibitemOpen
	\bibfield  {author} {\bibinfo {author} {\bibfnamefont {H.-K.}\ \bibnamefont
			{Lau}}, \bibinfo {author} {\bibfnamefont {R.}~\bibnamefont {Pooser}},
		\bibinfo {author} {\bibfnamefont {G.}~\bibnamefont {Siopsis}}, \ and\
		\bibinfo {author} {\bibfnamefont {C.}~\bibnamefont {Weedbrook}},\ }\href
	{\doibase 10.1103/PhysRevLett.118.080501} {\bibfield  {journal} {\bibinfo
			{journal} {Phys. Rev. Lett.}\ }\textbf {\bibinfo {volume} {118}},\ \bibinfo
		{pages} {080501} (\bibinfo {year} {2017})}\BibitemShut {NoStop}%
	\bibitem [{\citenamefont {Kempe}(2003)}]{Kempe2008}%
	\BibitemOpen
	\bibfield  {author} {\bibinfo {author} {\bibfnamefont {J.}~\bibnamefont
			{Kempe}},\ }\href {\doibase 10.1080/00107151031000110776} {\bibfield
		{journal} {\bibinfo  {journal} {Contemporary Physics}\ }\textbf {\bibinfo
			{volume} {44}},\ \bibinfo {pages} {307} (\bibinfo {year} {2003})}\BibitemShut
	{NoStop}%
	\bibitem [{\citenamefont {Dervovic}\ \emph {et~al.}(2018)\citenamefont
		{Dervovic}, \citenamefont {Herbster}, \citenamefont {Mountney}, \citenamefont
		{Severini}, \citenamefont {Usher},\ and\ \citenamefont
		{Wossnig}}]{Deervovic2018}%
	\BibitemOpen
	\bibfield  {author} {\bibinfo {author} {\bibfnamefont {D.}~\bibnamefont
			{Dervovic}}, \bibinfo {author} {\bibfnamefont {M.}~\bibnamefont {Herbster}},
		\bibinfo {author} {\bibfnamefont {P.}~\bibnamefont {Mountney}}, \bibinfo
		{author} {\bibfnamefont {S.}~\bibnamefont {Severini}}, \bibinfo {author}
		{\bibfnamefont {N.}~\bibnamefont {Usher}}, \ and\ \bibinfo {author}
		{\bibfnamefont {L.}~\bibnamefont {Wossnig}},\ }\href@noop {} {\enquote
		{\bibinfo {title} {Quantum linear systems algorithms: a primer},}\ }
	(\bibinfo {year} {2018}),\ \bibinfo {note} {arXiv:1802.08227}\BibitemShut
	{NoStop}%
	\bibitem [{\citenamefont {Childs}\ \emph {et~al.}(2003)\citenamefont {Childs},
		\citenamefont {Cleve}, \citenamefont {Deotto}, \citenamefont {Farhi},
		\citenamefont {Gutmann},\ and\ \citenamefont {Spielman}}]{Childs2003}%
	\BibitemOpen
	\bibfield  {author} {\bibinfo {author} {\bibfnamefont {A.~M.}\ \bibnamefont
			{Childs}}, \bibinfo {author} {\bibfnamefont {R.}~\bibnamefont {Cleve}},
		\bibinfo {author} {\bibfnamefont {E.}~\bibnamefont {Deotto}}, \bibinfo
		{author} {\bibfnamefont {E.}~\bibnamefont {Farhi}}, \bibinfo {author}
		{\bibfnamefont {S.}~\bibnamefont {Gutmann}}, \ and\ \bibinfo {author}
		{\bibfnamefont {D.~A.}\ \bibnamefont {Spielman}},\ }in\ \href {\doibase
		10.1145/780542.780552} {\emph {\bibinfo {booktitle} {Proceedings of the
				Thirty-Fifth Annual ACM Symposium on Theory of Computing}}},\ \bibinfo
	{series and number} {STOC ’03}\ (\bibinfo  {publisher} {Association for
		Computing Machinery},\ \bibinfo {address} {New York, NY, USA},\ \bibinfo
	{year} {2003})\ p.\ \bibinfo {pages} {59–68}\BibitemShut {NoStop}%
	\bibitem [{\citenamefont {Marvian}\ and\ \citenamefont
		{Lloyd}(2016)}]{Marvian2016}%
	\BibitemOpen
	\bibfield  {author} {\bibinfo {author} {\bibfnamefont {I.}~\bibnamefont
			{Marvian}}\ and\ \bibinfo {author} {\bibfnamefont {S.}~\bibnamefont
			{Lloyd}},\ }\href@noop {} {\enquote {\bibinfo {title} {Universal quantum
				emulator},}\ } (\bibinfo {year} {2016}),\ \bibinfo {note}
	{arXiv:1606.02734}\BibitemShut {NoStop}%
	\bibitem [{\citenamefont {Krantz}\ \emph {et~al.}(2019)\citenamefont {Krantz},
		\citenamefont {Kjaergaard}, \citenamefont {Yan}, \citenamefont {Orlando},
		\citenamefont {Gustavsson},\ and\ \citenamefont {Oliver}}]{Krantz2019}%
	\BibitemOpen
	\bibfield  {author} {\bibinfo {author} {\bibfnamefont {P.}~\bibnamefont
			{Krantz}}, \bibinfo {author} {\bibfnamefont {M.}~\bibnamefont {Kjaergaard}},
		\bibinfo {author} {\bibfnamefont {F.}~\bibnamefont {Yan}}, \bibinfo {author}
		{\bibfnamefont {T.~P.}\ \bibnamefont {Orlando}}, \bibinfo {author}
		{\bibfnamefont {S.}~\bibnamefont {Gustavsson}}, \ and\ \bibinfo {author}
		{\bibfnamefont {W.~D.}\ \bibnamefont {Oliver}},\ }\href {\doibase
		10.1063/1.5089550} {\bibfield  {journal} {\bibinfo  {journal} {Applied
				Physics Reviews}\ }\textbf {\bibinfo {volume} {6}},\ \bibinfo {pages}
		{021318} (\bibinfo {year} {2019})}\BibitemShut {NoStop}%
	\bibitem [{\citenamefont {Nielsen}\ and\ \citenamefont
		{Chuang}(2010)}]{Nielsen2010}%
	\BibitemOpen
	\bibfield  {author} {\bibinfo {author} {\bibfnamefont {M.~A.}\ \bibnamefont
			{Nielsen}}\ and\ \bibinfo {author} {\bibfnamefont {I.~L.}\ \bibnamefont
			{Chuang}},\ }\href {\doibase 10.1017/CBO9780511976667} {\emph {\bibinfo
			{title} {Quantum Computation and Quantum Information: 10th Anniversary
				Edition}}}\ (\bibinfo  {publisher} {Cambridge University Press},\ \bibinfo
	{year} {2010})\BibitemShut {NoStop}%
	\bibitem [{\citenamefont {Horodecki}\ \emph {et~al.}(1999)\citenamefont
		{Horodecki}, \citenamefont {Horodecki},\ and\ \citenamefont
		{Horodecki}}]{Horodecki1999}%
	\BibitemOpen
	\bibfield  {author} {\bibinfo {author} {\bibfnamefont {M.}~\bibnamefont
			{Horodecki}}, \bibinfo {author} {\bibfnamefont {P.}~\bibnamefont
			{Horodecki}}, \ and\ \bibinfo {author} {\bibfnamefont {R.}~\bibnamefont
			{Horodecki}},\ }\href {\doibase 10.1103/PhysRevA.60.1888} {\bibfield
		{journal} {\bibinfo  {journal} {Phys. Rev. A}\ }\textbf {\bibinfo {volume}
			{60}},\ \bibinfo {pages} {1888} (\bibinfo {year} {1999})}\BibitemShut
	{NoStop}%
	\bibitem [{\citenamefont {Schumacher}(1996)}]{Schumacher1996}%
	\BibitemOpen
	\bibfield  {author} {\bibinfo {author} {\bibfnamefont {B.}~\bibnamefont
			{Schumacher}},\ }\href {\doibase 10.1103/PhysRevA.54.2614} {\bibfield
		{journal} {\bibinfo  {journal} {Phys. Rev. A}\ }\textbf {\bibinfo {volume}
			{54}},\ \bibinfo {pages} {2614} (\bibinfo {year} {1996})}\BibitemShut
	{NoStop}%
	\bibitem [{\citenamefont {Johansson}\ \emph {et~al.}(2012)\citenamefont
		{Johansson}, \citenamefont {Nation},\ and\ \citenamefont {Nori}}]{qutip}%
	\BibitemOpen
	\bibfield  {author} {\bibinfo {author} {\bibfnamefont {J.}~\bibnamefont
			{Johansson}}, \bibinfo {author} {\bibfnamefont {P.}~\bibnamefont {Nation}}, \
		and\ \bibinfo {author} {\bibfnamefont {F.}~\bibnamefont {Nori}},\ }\href
	{\doibase https://doi.org/10.1016/j.cpc.2012.02.021} {\bibfield  {journal}
		{\bibinfo  {journal} {Computer Physics Communications}\ }\textbf {\bibinfo
			{volume} {183}},\ \bibinfo {pages} {1760 } (\bibinfo {year}
		{2012})}\BibitemShut {NoStop}%
	\bibitem [{\citenamefont {Wendin}(2017)}]{Wendin2017}%
	\BibitemOpen
	\bibfield  {author} {\bibinfo {author} {\bibfnamefont {G.}~\bibnamefont
			{Wendin}},\ }\href {\doibase 10.1088/1361-6633/aa7e1a} {\bibfield  {journal}
		{\bibinfo  {journal} {Reports on Progress in Physics}\ }\textbf {\bibinfo
			{volume} {80}},\ \bibinfo {pages} {106001} (\bibinfo {year}
		{2017})}\BibitemShut {NoStop}%
	\bibitem [{\citenamefont {Jin}\ \emph {et~al.}(2015)\citenamefont {Jin},
		\citenamefont {Kamal}, \citenamefont {Sears}, \citenamefont {Gudmundsen},
		\citenamefont {Hover}, \citenamefont {Miloshi}, \citenamefont {Slattery},
		\citenamefont {Yan}, \citenamefont {Yoder}, \citenamefont {Orlando},
		\citenamefont {Gustavsson},\ and\ \citenamefont {Oliver}}]{Jin2015}%
	\BibitemOpen
	\bibfield  {author} {\bibinfo {author} {\bibfnamefont {X.~Y.}\ \bibnamefont
			{Jin}}, \bibinfo {author} {\bibfnamefont {A.}~\bibnamefont {Kamal}}, \bibinfo
		{author} {\bibfnamefont {A.~P.}\ \bibnamefont {Sears}}, \bibinfo {author}
		{\bibfnamefont {T.}~\bibnamefont {Gudmundsen}}, \bibinfo {author}
		{\bibfnamefont {D.}~\bibnamefont {Hover}}, \bibinfo {author} {\bibfnamefont
			{J.}~\bibnamefont {Miloshi}}, \bibinfo {author} {\bibfnamefont
			{R.}~\bibnamefont {Slattery}}, \bibinfo {author} {\bibfnamefont
			{F.}~\bibnamefont {Yan}}, \bibinfo {author} {\bibfnamefont {J.}~\bibnamefont
			{Yoder}}, \bibinfo {author} {\bibfnamefont {T.~P.}\ \bibnamefont {Orlando}},
		\bibinfo {author} {\bibfnamefont {S.}~\bibnamefont {Gustavsson}}, \ and\
		\bibinfo {author} {\bibfnamefont {W.~D.}\ \bibnamefont {Oliver}},\ }\href
	{\doibase 10.1103/PhysRevLett.114.240501} {\bibfield  {journal} {\bibinfo
			{journal} {Phys. Rev. Lett.}\ }\textbf {\bibinfo {volume} {114}},\ \bibinfo
		{pages} {240501} (\bibinfo {year} {2015})}\BibitemShut {NoStop}%
	\bibitem [{\citenamefont {Koch}\ \emph {et~al.}(2007)\citenamefont {Koch},
		\citenamefont {Yu}, \citenamefont {Gambetta}, \citenamefont {Houck},
		\citenamefont {Schuster}, \citenamefont {Majer}, \citenamefont {Blais},
		\citenamefont {Devoret}, \citenamefont {Girvin},\ and\ \citenamefont
		{Schoelkopf}}]{Koch2007}%
	\BibitemOpen
	\bibfield  {author} {\bibinfo {author} {\bibfnamefont {J.}~\bibnamefont
			{Koch}}, \bibinfo {author} {\bibfnamefont {T.~M.}\ \bibnamefont {Yu}},
		\bibinfo {author} {\bibfnamefont {J.}~\bibnamefont {Gambetta}}, \bibinfo
		{author} {\bibfnamefont {A.~A.}\ \bibnamefont {Houck}}, \bibinfo {author}
		{\bibfnamefont {D.~I.}\ \bibnamefont {Schuster}}, \bibinfo {author}
		{\bibfnamefont {J.}~\bibnamefont {Majer}}, \bibinfo {author} {\bibfnamefont
			{A.}~\bibnamefont {Blais}}, \bibinfo {author} {\bibfnamefont {M.~H.}\
			\bibnamefont {Devoret}}, \bibinfo {author} {\bibfnamefont {S.~M.}\
			\bibnamefont {Girvin}}, \ and\ \bibinfo {author} {\bibfnamefont {R.~J.}\
			\bibnamefont {Schoelkopf}},\ }\href {\doibase 10.1103/PhysRevA.76.042319}
	{\bibfield  {journal} {\bibinfo  {journal} {Phys. Rev. A}\ }\textbf {\bibinfo
			{volume} {76}},\ \bibinfo {pages} {042319} (\bibinfo {year}
		{2007})}\BibitemShut {NoStop}%
	\bibitem [{\citenamefont {Schreier}\ \emph {et~al.}(2008)\citenamefont
		{Schreier}, \citenamefont {Houck}, \citenamefont {Koch}, \citenamefont
		{Schuster}, \citenamefont {Johnson}, \citenamefont {Chow}, \citenamefont
		{Gambetta}, \citenamefont {Majer}, \citenamefont {Frunzio}, \citenamefont
		{Devoret}, \citenamefont {Girvin},\ and\ \citenamefont
		{Schoelkopf}}]{Schreier2008}%
	\BibitemOpen
	\bibfield  {author} {\bibinfo {author} {\bibfnamefont {J.~A.}\ \bibnamefont
			{Schreier}}, \bibinfo {author} {\bibfnamefont {A.~A.}\ \bibnamefont {Houck}},
		\bibinfo {author} {\bibfnamefont {J.}~\bibnamefont {Koch}}, \bibinfo {author}
		{\bibfnamefont {D.~I.}\ \bibnamefont {Schuster}}, \bibinfo {author}
		{\bibfnamefont {B.~R.}\ \bibnamefont {Johnson}}, \bibinfo {author}
		{\bibfnamefont {J.~M.}\ \bibnamefont {Chow}}, \bibinfo {author}
		{\bibfnamefont {J.~M.}\ \bibnamefont {Gambetta}}, \bibinfo {author}
		{\bibfnamefont {J.}~\bibnamefont {Majer}}, \bibinfo {author} {\bibfnamefont
			{L.}~\bibnamefont {Frunzio}}, \bibinfo {author} {\bibfnamefont {M.~H.}\
			\bibnamefont {Devoret}}, \bibinfo {author} {\bibfnamefont {S.~M.}\
			\bibnamefont {Girvin}}, \ and\ \bibinfo {author} {\bibfnamefont {R.~J.}\
			\bibnamefont {Schoelkopf}},\ }\href {\doibase 10.1103/PhysRevB.77.180502}
	{\bibfield  {journal} {\bibinfo  {journal} {Phys. Rev. B}\ }\textbf {\bibinfo
			{volume} {77}},\ \bibinfo {pages} {180502(R)} (\bibinfo {year}
		{2008})}\BibitemShut {NoStop}%
	\bibitem [{\citenamefont {Kounalakis}\ \emph {et~al.}(2018)\citenamefont
		{Kounalakis}, \citenamefont {Dickel}, \citenamefont {Bruno}, \citenamefont
		{Langford},\ and\ \citenamefont {Steele}}]{Kounalakis2018}%
	\BibitemOpen
	\bibfield  {author} {\bibinfo {author} {\bibfnamefont {M.}~\bibnamefont
			{Kounalakis}}, \bibinfo {author} {\bibfnamefont {C.}~\bibnamefont {Dickel}},
		\bibinfo {author} {\bibfnamefont {A.}~\bibnamefont {Bruno}}, \bibinfo
		{author} {\bibfnamefont {N.~K.}\ \bibnamefont {Langford}}, \ and\ \bibinfo
		{author} {\bibfnamefont {G.~A.}\ \bibnamefont {Steele}},\ }\href {\doibase
		10.1038/s41534-018-0088-9} {\bibfield  {journal} {\bibinfo  {journal} {npj
				Quantum Information}\ }\textbf {\bibinfo {volume} {4}},\ \bibinfo {pages}
		{38} (\bibinfo {year} {2018})}\BibitemShut {NoStop}%
	\bibitem [{\citenamefont {Zhao}\ \emph {et~al.}(2020)\citenamefont {Zhao},
		\citenamefont {Xu}, \citenamefont {Lan}, \citenamefont {Tan}, \citenamefont
		{Yu},\ and\ \citenamefont {Yu}}]{Zhao2020}%
	\BibitemOpen
	\bibfield  {author} {\bibinfo {author} {\bibfnamefont {P.}~\bibnamefont
			{Zhao}}, \bibinfo {author} {\bibfnamefont {P.}~\bibnamefont {Xu}}, \bibinfo
		{author} {\bibfnamefont {D.}~\bibnamefont {Lan}}, \bibinfo {author}
		{\bibfnamefont {X.}~\bibnamefont {Tan}}, \bibinfo {author} {\bibfnamefont
			{H.}~\bibnamefont {Yu}}, \ and\ \bibinfo {author} {\bibfnamefont
			{Y.}~\bibnamefont {Yu}},\ }\href@noop {} {\enquote {\bibinfo {title}
			{High-contrast zz interaction using multi-type superconducting qubits},}\ }
	(\bibinfo {year} {2020}),\ \bibinfo {note} {arXiv:2002.07560}\BibitemShut
	{NoStop}%
	\bibitem [{\citenamefont {Chen}\ \emph {et~al.}(2014)\citenamefont {Chen},
		\citenamefont {Neill}, \citenamefont {Roushan}, \citenamefont {Leung},
		\citenamefont {Fang}, \citenamefont {Barends}, \citenamefont {Kelly},
		\citenamefont {Campbell}, \citenamefont {Chen}, \citenamefont {Chiaro},
		\citenamefont {Dunsworth}, \citenamefont {Jeffrey}, \citenamefont {Megrant},
		\citenamefont {Mutus}, \citenamefont {O'Malley}, \citenamefont {Quintana},
		\citenamefont {Sank}, \citenamefont {Vainsencher}, \citenamefont {Wenner},
		\citenamefont {White}, \citenamefont {Geller}, \citenamefont {Cleland},\ and\
		\citenamefont {Martinis}}]{Chen2014}%
	\BibitemOpen
	\bibfield  {author} {\bibinfo {author} {\bibfnamefont {Y.}~\bibnamefont
			{Chen}}, \bibinfo {author} {\bibfnamefont {C.}~\bibnamefont {Neill}},
		\bibinfo {author} {\bibfnamefont {P.}~\bibnamefont {Roushan}}, \bibinfo
		{author} {\bibfnamefont {N.}~\bibnamefont {Leung}}, \bibinfo {author}
		{\bibfnamefont {M.}~\bibnamefont {Fang}}, \bibinfo {author} {\bibfnamefont
			{R.}~\bibnamefont {Barends}}, \bibinfo {author} {\bibfnamefont
			{J.}~\bibnamefont {Kelly}}, \bibinfo {author} {\bibfnamefont
			{B.}~\bibnamefont {Campbell}}, \bibinfo {author} {\bibfnamefont
			{Z.}~\bibnamefont {Chen}}, \bibinfo {author} {\bibfnamefont {B.}~\bibnamefont
			{Chiaro}}, \bibinfo {author} {\bibfnamefont {A.}~\bibnamefont {Dunsworth}},
		\bibinfo {author} {\bibfnamefont {E.}~\bibnamefont {Jeffrey}}, \bibinfo
		{author} {\bibfnamefont {A.}~\bibnamefont {Megrant}}, \bibinfo {author}
		{\bibfnamefont {J.~Y.}\ \bibnamefont {Mutus}}, \bibinfo {author}
		{\bibfnamefont {P.~J.~J.}\ \bibnamefont {O'Malley}}, \bibinfo {author}
		{\bibfnamefont {C.~M.}\ \bibnamefont {Quintana}}, \bibinfo {author}
		{\bibfnamefont {D.}~\bibnamefont {Sank}}, \bibinfo {author} {\bibfnamefont
			{A.}~\bibnamefont {Vainsencher}}, \bibinfo {author} {\bibfnamefont
			{J.}~\bibnamefont {Wenner}}, \bibinfo {author} {\bibfnamefont {T.~C.}\
			\bibnamefont {White}}, \bibinfo {author} {\bibfnamefont {M.~R.}\ \bibnamefont
			{Geller}}, \bibinfo {author} {\bibfnamefont {A.~N.}\ \bibnamefont {Cleland}},
		\ and\ \bibinfo {author} {\bibfnamefont {J.~M.}\ \bibnamefont {Martinis}},\
	}\href {\doibase 10.1103/PhysRevLett.113.220502} {\bibfield  {journal}
		{\bibinfo  {journal} {Phys. Rev. Lett.}\ }\textbf {\bibinfo {volume} {113}},\
		\bibinfo {pages} {220502} (\bibinfo {year} {2014})}\BibitemShut {NoStop}%
	\bibitem [{\citenamefont {D\"ur}\ \emph {et~al.}(2000)\citenamefont {D\"ur},
		\citenamefont {Vidal},\ and\ \citenamefont {Cirac}}]{Dur2000}%
	\BibitemOpen
	\bibfield  {author} {\bibinfo {author} {\bibfnamefont {W.}~\bibnamefont
			{D\"ur}}, \bibinfo {author} {\bibfnamefont {G.}~\bibnamefont {Vidal}}, \ and\
		\bibinfo {author} {\bibfnamefont {J.~I.}\ \bibnamefont {Cirac}},\ }\href
	{\doibase 10.1103/PhysRevA.62.062314} {\bibfield  {journal} {\bibinfo
			{journal} {Phys. Rev. A}\ }\textbf {\bibinfo {volume} {62}},\ \bibinfo
		{pages} {062314} (\bibinfo {year} {2000})}\BibitemShut {NoStop}%
	\bibitem [{\citenamefont {{Devoret}}(1997)}]{Devoret1997}%
	\BibitemOpen
	\bibfield  {author} {\bibinfo {author} {\bibfnamefont {M.~H.}\ \bibnamefont
			{{Devoret}}},\ }in\ \href@noop {} {\emph {\bibinfo {booktitle} {Fluctuations
				Quantiques/Quantum Fluctuations: Les Houches Session LXIII}}},\ \bibinfo
	{editor} {edited by\ \bibinfo {editor} {\bibfnamefont {S.}~\bibnamefont
			{{Reynaud}}}, \bibinfo {editor} {\bibfnamefont {E.}~\bibnamefont
			{{Giacobino}}}, \ and\ \bibinfo {editor} {\bibfnamefont {J.}~\bibnamefont
			{{Zinn-Justin}}}}\ (\bibinfo  {publisher} {Elsevier},\ \bibinfo {year}
	{1997})\ p.\ \bibinfo {pages} {351}\BibitemShut {NoStop}%
	\bibitem [{\citenamefont {Vool}\ and\ \citenamefont
		{Devoret}(2017)}]{Devoret2017}%
	\BibitemOpen
	\bibfield  {author} {\bibinfo {author} {\bibfnamefont {U.}~\bibnamefont
			{Vool}}\ and\ \bibinfo {author} {\bibfnamefont {M.~H.}\ \bibnamefont
			{Devoret}},\ }\href@noop {} {\bibfield  {journal} {\bibinfo  {journal}
			{International Journal of Circuit Theory and Applications}\ }\textbf
		{\bibinfo {volume} {45}},\ \bibinfo {pages} {897} (\bibinfo {year}
		{2017})}\BibitemShut {NoStop}%
\end{thebibliography}
%

\clearpage
\onecolumngrid
\appendix

\section{Analysis of the superconducting circuit}\label{app:analCircuit}

\begin{figure}
	\centering
	\includegraphics[scale=0.5]{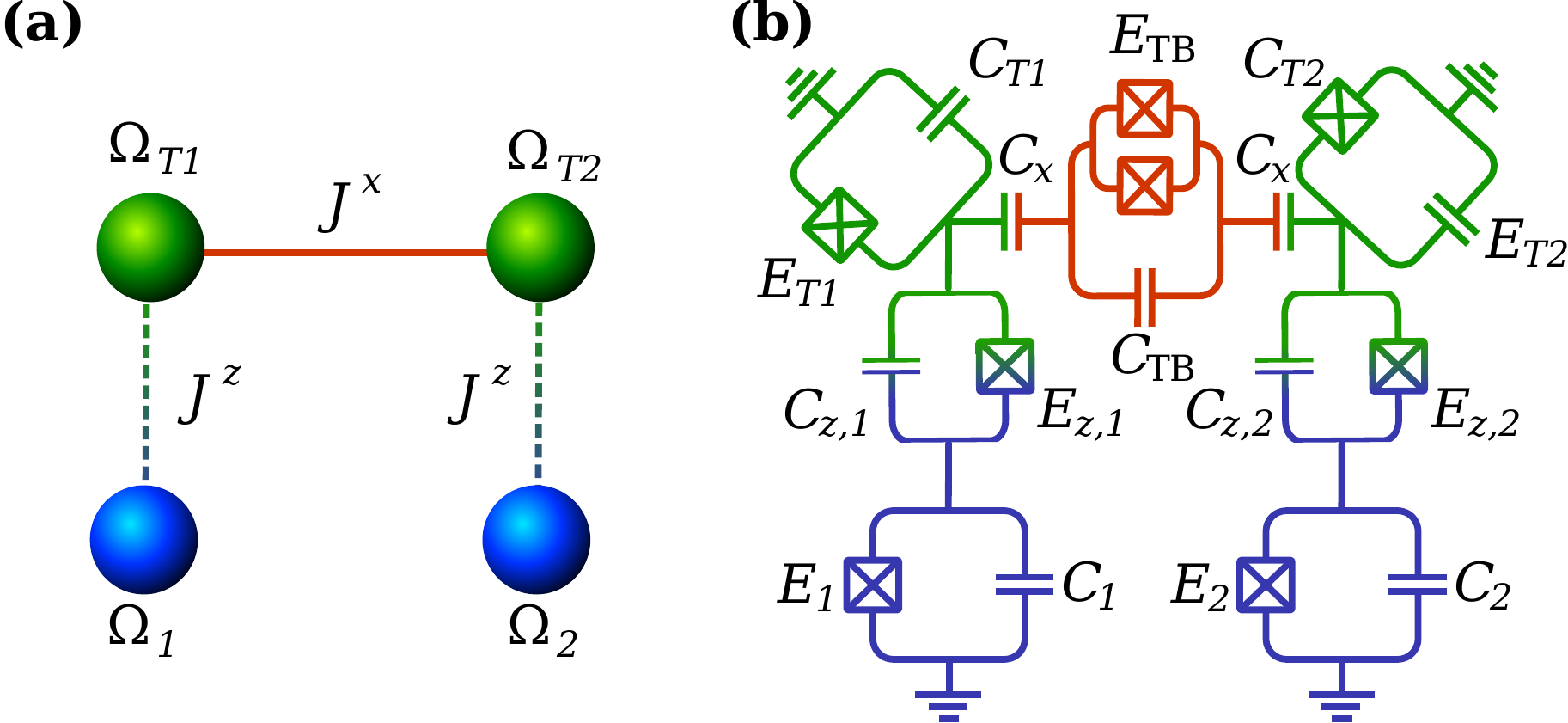}
	\caption{Implementation of the \textsc{c}$^2$\iswap gate. Figure \textbf{(a)} shows a schematic representation of the model implementing the \textsc{c}$^2$\iswap gate, with the green spheres (subscript $T1$ and $T2$) representing the target qubit and the blue spheres (subscript 1 and 2) representing the control qubits. Figure \textbf{(b)} shows the superconducting circuit yielding the model in figure \textbf{(a)}. The different parts of the system are colored according to their role, as per \textbf{(a)}. Note that in this implementation each target qubit are connected to a control qubit, contrary to the proposed implementation where both control qubits are connected to the first target qubit. It is, however, irrelevant which qubit the control qubits are connected to, the controlled \iswap gate can be created either way.}
	\label{fig:cfredkin}
\end{figure}

We consider two fixed frequency target qubits connected by a tunable bus qubit as in Ref. \cite{McKay2016}. We further connect a number, $n$, of fixed frequency qubits to the target qubits. It is irrelevant which target qubit these control qubits are connected to. The control qubits can be connected to either (or both) target qubits, but for simplicity we connect all control qubits to target qubit 1. An example of the circuit with one control qubit can be seen in \cref{fig:fredkin}(b), while an example of the case of $n=2$ qubits can be seen in \cref{fig:cfredkin}(b).
Following the procedure of Refs. \cite{Devoret1997,Devoret2017} we obtain the following Lagrangian 
\begin{equation}\label{eq:Lagrangian}
\begin{aligned}
L =& 2 \sum_{\substack{i=T1,T2, \\ TB,TB',1,...}}^{n} \left[C_i\dot\varphi_i^2 + E_i\cos\varphi_i \right] + 2C_x\left( \dot \varphi_{T1} - \dot \varphi_{TB} \right)^2 + 2C_x\left( \dot \varphi_{T2} - \dot \varphi_\text{TB'} \right)^2  + 2C_{TB}(\dot\varphi_{TB} - \dot\varphi_\text{TB'})^2\\
&+ 2 \sum_{i=1}^{n} C_{z,i}\left(\dot\varphi_i - \dot \varphi_{T1} \right)^2   +  \sum_{i=1}^n E_{z,i}\cos(\varphi_{T1}-\varphi_i)+ E_{TB} \left[\cos(\varphi_{TB}-\varphi_\text{TB'} + \Phi) + \cos(\varphi_{TB}-\varphi_\text{TB'} + \Phi)\right],
\end{aligned}
\end{equation}
where the first summation is understood as the summation over $T1,T2,1,2,\dots,n$. We denote $\varphi_{T1}$ and $\varphi_{T2}$ as the node fluxes of the target qubits (green islands in \cref{fig:fredkin,fig:cfredkin}(b)), $\varphi_{TB}$ and $\varphi_\text{TB'}$ are the node fluxes of the tunable bus (red islands in \cref{fig:fredkin,fig:cfredkin}(b)), and $\varphi_i$ are the node fluxes of the control qubits (blue islands in \cref{fig:fredkin,fig:cfredkin}(b)). $\Phi$ is the external magnetic flux through the tunable bus in units of $\Phi_0/2\pi$, where $\Phi_0$ is the flux quantum.

Using Kirchoff's rules we find that $\varphi_{TB'} = \Phi - \varphi_{TB}$ which we use to rewrite the last term of \cref{eq:Lagrangian} to
\begin{equation}
	\cos(\varphi_{TB}-\varphi_\text{TB'} + \Phi) + \cos(\varphi_{TB}-\varphi_\text{TB'} + \Phi) = 2  \cos(2\Phi) \cos (2\varphi_{TB}).
\end{equation}
With this the Lagrangian becomes
\begin{equation}
\begin{aligned}
L =& 2 \sum_{\substack{i=T1,T2, \\ TB,TB',1,...}}^{n} \left[C_i\dot\varphi_i^2 + E_i\cos\varphi_i \right] + 2C_x\left( \dot \varphi_{T1}^2 + \dot\varphi_{T2}^2 + 2\dot \varphi_{TB}^2 + 2\dot\varphi_{TB}\left( \dot \varphi_{T2} - \dot \varphi_{T1} \right)\right)  + 4C_{TB}\dot\varphi_{TB}^2 \\
&+ 2 \sum_{i=1}^{n} C_{z,i}\left(\dot\varphi_i - \dot \varphi_{T1} \right)^2   +  \sum_{i=1}^n E_{z,i}\cos(\varphi_{T1}-\varphi_i)+ 2 E_{TB} \cos(2\Phi) \cos (2\varphi_{TB}),
\end{aligned}
\end{equation}
where we have ignored all terms concerning $\dot\Phi$ since these all contribute with irrelevant constant terms.

The terms coming from the capacitors and are interpreted as kinetic terms, while the remaining terms come from the Josephson junctions and are interpreted as potential terms. The $n$ indicates the number of blue islands on the circuit diagram, i.e., for the \textsc{c}\iswap in \cref{fig:fredkin}(b) $n=1$. Considering the case of a single control qubit and arranging the node fluxes in a vector as $\vec{\varphi}^T = (\varphi_1,\varphi_{T1},\varphi_{TB},\varphi_{T2})$, the capacitance matrix becomes
\begin{equation}
K = \begin{bmatrix}
C_1 + C_{z,1} & -C_{z,1} & 0 & 0 \\
-C_{z,1} & C_{T1}+C_{z,1}+2C_x & -C_x & 0\\
0 & -C_x &  4C_{TB}+2C_x & -C_x \\
0 & 0 & -C_x & C_{T2} + 2C_x
\end{bmatrix}.
\end{equation}
For two control qubits (see \cref{fig:cfredkin}(b) for circuit diagram of this gate) the capacitance matrix takes the form
\begin{equation}
K = \begin{bmatrix}
C_1 + C_{z,1} & 0 & -C_{z,1} & 0 & 0 \\
0 & C_2 + C_{z,2} & -C_{z,2} & 0 & 0 \\
-C_{z,1} & -C_{z,2} & C_{T1}+C_{z,1}+2C_x & -C_x & 0\\
0 & 0 & -C_x & 4C_{TB} +2C_x & -C_x \\
0 & 0 & 0 & -C_x & C_{T2} + 2C_x
\end{bmatrix}.
\end{equation}
and so on for higher $n$. We this we can do a Legendre transform and write the Hamiltonian as
\begin{equation}
	H = \frac{1}{2} \hat{\vec{p}}^T K^{-1} \hat{\vec{p}} + U(\vec \varphi),
\end{equation}
where $\vec p^T = (p_1, p_{T1}, p_{TB}, p_{T2})$ is the conjugate momentum and $U(\vec \varphi)$ is the potential energy coming from the Josephson junctions.

The typical transmon have a charging energy much smaller than the junction energy, and therefore the phase is well localized near the bottom of the potential. We can therefore expand the potential part of the Hamiltonian to fourth order
\begin{equation*}
U(\varphi) = \sum_{i=T1,T2,1}^{n} E_i \left[\frac{1}{2}\varphi_i^2 - \frac{1}{24}\varphi^4_i\right]+ \sum_{i=1}^{n} E_{z,i} \left[\frac{1}{2}\left(\varphi_i - \varphi_{T1}\right)^2 - \frac{1}{24}\left(\varphi_i - \varphi_{T1}\right)^4\right] + 8 E_{TB} \cos(2\Phi) \left[  \frac{1}{2}\varphi_{TB}^2 - \frac{4}{24}\varphi^4_{TB}\right].
\end{equation*}
By collecting terms we can write the full Hamiltonian as
\begin{align*}
H=& \sum_{\substack{i=T1,T2,1}}^{n} \left[ \frac{1}{2}E^C_ip_i^2 + \frac{1}{2}E^J_i\varphi^2_i - \frac{1}{24}E^J_i\varphi_i^4 \right] +\frac{1}{2}\sum_{\substack{i\neq j = \\ T1,T2,TB,1}}^n(K^{-1})_{(i,j)}p_ip_{j} \\
& + \sum_{i=1}^{n}E_{z,i} \left[ - \frac{1}{4}\varphi_i^2\varphi_{T1}^2 -\varphi_i\varphi_{T1} + \frac{1}{6}\left(\varphi^3_i\varphi_{T1} + \varphi_i\varphi_{T1}^3\right) \right] + \frac{1}{2}E^C_{TB}p_{TB}^2 + \frac{1}{2}E^J_{TB}\varphi^2_{TB} - \frac{1}{6}E^J_{TB}\varphi_{TB}^4,
\end{align*}
where the effective energy of the capacitances is $E^C_i  = (K^{-1})_{(i,i)}/8$. The second summation is understood as the sum over $i$ and $j = T1,T2,TB,1,\dots,n$, where $i$ and $j$ is never equal. Note that there is a capacitive coupling between all of the qubits regardless of whether there actually is a capacitor between them. The effective Josephson energies are
\begin{subequations}
	\begin{align}
	E^J_i =& E_i + 2E_{z,i},\\
	E^J_{T1} =& E_{T1} + \sum_{i=1}^n E_{z,i}, \\
	E^J_{T2} =& E_{T2}, \\
	E^J_{TB} =& 8E_{TB}\cos(2\Phi).
	\end{align}
\end{subequations}
We now do the canonical quantization $\varphi_i \rightarrow \hat \varphi_i$ and $p_i \rightarrow \hat p_i$, requiring that $[\hat p_i,\hat \varphi_j] = \delta_{ij}$. This allows us to change into step operators 
\begin{equation}
\hat\varphi_i = \sqrt{\frac{\zeta_i}{2}}\left(\hat b_i^\dagger + \hat b_i\right), \qquad \hat p_i = \frac{i}{\sqrt{2\zeta_i}}\left(\hat b_i^\dagger - \hat b_i\right),
\end{equation}
with impedance $\zeta_i = \sqrt{(K^{-1})_{(i,i)}/E^J_i}$, the Hamiltonian takes the form
\begin{align*}
	\hat H=&  \sum_{i=T1,T2,1}^{n}\left[\sqrt{8E^C_iE^J_i} \hat b_i^\dagger \hat b_i - \frac{E^J_i}{12} \left(\hat b_i^\dagger + \hat b_i\right)^4\right] - \frac{1}{2}\sum_{\substack{i\neq j = \\ T1,T2,TB,1}}^n\frac{(K^{-1})_{(i,j)}}{\sqrt{\zeta_i\zeta_j}}\left(\hat b_i^\dagger - \hat b_i\right)\left(\hat b_j^\dagger - \hat b_j\right) \\
	&+\sum_{i=1}^{n} E_{z,i} \left[ - \frac{1}{24}\zeta_i\zeta_{T1}\left(\hat b_i^\dagger + \hat b_i\right)^2\left(\hat b_{T1}^\dagger + \hat b_{T1}\right)^2 -\frac{1}{2}\sqrt{\zeta_i\zeta_{T1}}\left(\hat b_i^\dagger + \hat b_i\right)\left(\hat b_{T1}^\dagger + \hat b_{T1}\right)   \right. \\
	&\phantom{+\sum_{i=1}^{n}}+ \left. \frac{1}{24}\left(\zeta_i\sqrt{\zeta_i\zeta_{T1}}\left(\hat b_i^\dagger + \hat b_i\right)^3\left(\hat b_{T1}^\dagger + \hat b_{T1}\right) + \zeta_{T1}\sqrt{\zeta_i\zeta_{T1}}\left(\hat b_i^\dagger + \hat b_i\right)\left(\hat b_{T1}^\dagger + \hat b_{T1}\right)^3\right) \right] \\
	& + \sqrt{8E^C_{TB}E^J_{TB}} \hat b_{TB}^\dagger \hat b_{TB} - \frac{1}{3}E^J_{TB} \left(\hat b_{TB}^\dagger + \hat b_{TB}\right)^4.
\end{align*}
If we operate the circuit in the weak coupling regime $E_{z,i} \ll E_j$ and $C_{z,i} \ll C_j$ for all $i$ and $j$ we can view the system as $n+2$ harmonic oscillators perturbed by their interactions. If we also assume that the modes of oscillator $T1$ and $T2$ are close to resonant, we can treat their detuning as part of the perturbation. The total Hamiltonian is then the sum of the harmonic oscillator Hamiltonian $\hat H_0$ and a perturbation $\hat V$. If we introduce the number operator $\hat n = \hat b^\dagger \hat b$ and swap operator $\hat X_{ij} = \hat b_j\hat b_i^\dagger +  \hat b_j ^\dagger \hat b_i$, we can write the two parts of the Hamiltonian as
\begin{subequations}
\begin{align}
	\hat H_0 =&  \sum_{i=1}^{n} \omega_i\hat n_i + \omega_{T2}(\hat n_{T1} + \hat n_{T2}) + \omega_{TB} \hat n_{TB}, \\
	\hat V =& \delta \hat n_{T1} - \frac{1}{2} \sum_{i=T1,T2,TB,1}^{n} E^C_i \hat n_i (\hat n_i - 1) + \sum_{i=1}^n   g_{iT1}^z \hat n_i \hat n_{T1} + \sum_{\substack{i\neq j = \\ T1,T2,TB,1}}^n g^x_{ij} \hat X_{ij}\\
	&+ \sum_{i=1}^n  g_{iT1}^{xz} \left[\zeta_i \left( \hat X_{iT1} \hat n_i + \hat n_i \hat X_{iT1}  \right) + \zeta_{T1}\left( \hat X_{iT1} \hat n_{T1} + \hat n_{T1} \hat X_{iT1}  \right)\right],  \nonumber
\end{align}
\end{subequations}
where the qubit frequencies are then given as
\begin{subequations}\label{eq:omega}
	\begin{align}
	\omega_i =& \sqrt{8E^J_iE_i^C} - \frac{1}{12} E_{z,i} \zeta_i\zeta_{T1} \qquad \text{for } i = 1,2,\dots n, \\
	\omega_{T1} =& \sqrt{8E^J_{T1}E_{T1}^C} - \frac{1}{12} E_{z,i} \zeta_{T1} \sum_{i=1}^n \zeta_{i}, \\
	\omega_{T2} =& \sqrt{8E^J_{T2}E_{T2}^C}, \\
	\omega_{TB} =& \sqrt{8E^J_{TB}E_{TB}^C},\\
	\delta =& \frac{1}{2}\left(\omega_{T1} - \omega_{T2}\right),
	\end{align}
\end{subequations}
and the coupling strengths are given as
\begin{subequations}
\begin{align}
	g_{iT1}^z =& - \frac{1}{4} E_{z,i} \zeta_i\zeta_{T1}, \qquad \text{for } i = 1,2,\dots,n, \\
	g_{iT1}^{xz} =& - \frac{1}{16} E_{z,i}\sqrt{\zeta_i\zeta_{T1}},\qquad \text{for }  i = 1,2,\dots,n,\\
	g_{ij}^x =& - \frac{1}{2} \frac{(K^{-1})_{(i,j)}}{\sqrt{\zeta_i\zeta_j}},\qquad \text{for } j,i = 1,2,\dots,n \label{eq:gijx},\\
	g_{TjTB}^x =& - \frac{1}{2} \frac{(K^{-1})_{(Tj,TB)}}{\sqrt{\zeta_{Tj}\zeta_{TB}}},\qquad \text{for } j = 1,2 \label{eq:gTiTBx},\\
	g_{iT1}^x =& - \frac{1}{2} \frac{(K^{-1})_{(i,T1)}}{\sqrt{\zeta_i\zeta_{T1}}} - \frac{1}{2}E_{z,i} \sqrt{\zeta_i\zeta_{T1}} + \frac{1}{16}E_{z,i}(\zeta_i + \zeta_{T1})\sqrt{\zeta_i\zeta_{T1}} ,\qquad \text{for } i = 1,2,\dots,n. \label{eq:giTjx}
\end{align}
\end{subequations}
If we only consider the two lowest lying states of each oscillator, the uncoupled Hamiltonian has a degenerate spectrum with $2^{n+2}$ states. If we require the detunings $\Delta_{ij} = \omega_i - \omega_{j}$ between each of the control qubits to be much larger than the transversal couplings in \cref{eq:gijx}, we can ignore first order excitations swaps between the control qubits. If we further require that the control qubits are detuned from the target qubits in such a way that $\Delta_{iTj} = \omega_i - \omega_{Tj}$ is much larger than the transversal coupling in \cref{eq:giTjx} we can also neglect first order excitation swaps between the target qubits and the control qubits. This leaves only one transversal coupling in \cref{eq:gTiTBx}.

If the anharmonicity is sufficiently larger than the transversal coupling between the target qubits we can justify projecting the final effective Hamiltonian into the two lowest states of each qubit. This projection is done using degenerate second order perturbation theory
In this case each degenerate subspace is well described by an effective interaction
\begin{equation}\label{eq:eff_ham_general_formula}
\hat{P}\hat{V}_\text{eff}\hat{P} = \hat{P}\hat{V}\hat{P}+\hat{P}\hat{V}\hat{Q}\frac{1}{E_D-\hat{Q}\hat{H}_0\hat{Q}}\hat{Q}\hat{V}\hat{P},
\end{equation}
where $\hat{P}$ projects onto the degenerate subspace consisting of the $2^{n+2}$ lowest lying states and $\hat{Q}=1-\hat{P}$ projects onto the orthogonal complement. Doing so yields an effective interaction between the qubits given by
\begin{equation}
\hat V_\text{eff} =-\frac{\Delta_{T1}}{2} \sigma_{T1}^z + \sum_{i=1}^n \frac{J_{i}^z}{2}\sigma^z_{T1}\sigma_i^z + \sum_{j=1}^2 g^x_{TjTB}\left(\sigma^+_{Tj}\sigma^-_{TB} + \sigma^-_{Tj}\sigma^+_{TB}\right).
\end{equation}
The detuning of the target qubit can then be calculated and the second order matrix elements are
\begin{equation}
\Delta_{T1} = -\delta + \sum_{ i=1}^n \left[\frac{g_{iT1}^z}{2} - \frac{g^x_{iT1} - g^{xz}_{iT1}(\zeta_i + 2\zeta_{T1})}{\Delta_{iT}}\right],
\end{equation}
where $\Delta_{iT}=\omega_i - \omega_{T2}$ is the detuning of the target qubits with respect to the $i$th control qubit. 
The longitudinal coupling between the target qubits and the control qubits are 
\begin{equation}\label{eq:Jz}
J^z_{i} = \frac{g^z_{iT1}}{2} + \frac{g^{xz}_{iT1}(\zeta_i - 2\zeta_{Tj})}{\Delta_{iT}}.
\end{equation}
As described in the main text, the purpose of this longitudinal coupling is to tune the target qubits in and out of resonance, depending on the state of the control qubits. We thus require this coupling to be significantly larger than the coupling between the target qubits and the tunable bus, $g^x_{TjTB}$.

We now wish to perform the same trick as Ref. \cite{McKay2016} in order to gain control over the transversal couplings to the target qubits. We there consider the dispersive regime, where $|g^x_{TjTB}/(\omega_{Tj} - \omega_{TB})| \ll 1$. In this regime we can adiabatically eliminate the tunable bus, which yields the following terms in the Hamiltonian
\begin{subequations}
\begin{align}
	\hat H_0 =& \sum_{i=1}^{n} \frac{\omega_i}{2} \sigma^z_i + \frac{\tilde \omega_{T2}}{2} (\sigma^z_{T1} + \sigma^z_{T2}), \\
	\hat V_\text{eff} =& -\frac{\tilde\Delta_{T1}}{2}\sigma_{T1}^z + \sum_{i=1}^n \frac{J_{i}^z}{2}\sigma^z_{T1}\sigma_i^z + \tilde J^x\left(\sigma^+_{T1}\sigma^-_{T2} + \sigma^-_{T1}\sigma^+_{T2}\right),
\end{align}
\end{subequations}
where the dressed qubit frequency and dressed detuning is
\begin{subequations}
\begin{align}
	\tilde \omega_{Tj} =& \omega_{Tj} + \frac{(g^x_{TjTB})^2}{\omega_{Tj} - \omega_{TB}},\\
	\tilde J^x =&\frac{g^x_{T1TB}g^x_{T2TB}}{2} \left( \frac{1}{\omega_{T1} - \omega_{TB}} + \frac{1}{\omega_{T2} - \omega_{TB}} \right).\label{eq:Jx}
\end{align}
\end{subequations}
In order to interact the qubits via the tunable bus coupler we apply a sinusoidal fast-flux bias modulation of amplitude, $\chi$, such that the flux applied to the tunable bus becomes $\Phi(t) = \Theta + \chi \cos (\omega_\Phi t)$. By expanding the dressed qubit frequency $\tilde\omega_{Tj}$ in the parameter $\chi \cos (\omega_\Phi t)$, where $\chi\ll 1$, we obtain
\begin{equation}
\begin{aligned}
	\tilde\omega_{Tj}(\Phi(t)) \simeq&\; \tilde \omega_{Tj} (\Theta) + \left. \frac{\partial \tilde \omega_{Tj}}{\partial \Phi} \right|_{\Phi \rightarrow \Theta} \chi \cos (\omega_\Phi t) +  \frac{1}{2}\left. \frac{\partial^2 \tilde \omega_{Tj}}{\partial \Phi^2} \right|_{\Phi \rightarrow \Theta} (\chi \cos (\omega_\Phi t))^2 \\
	=& \left[ \tilde \omega_{Tj} (\Theta) - \frac{\chi^2}{4} \left. \frac{\partial^2 \tilde \omega_{Tj}}{\partial \Phi^2} \right|_{\Phi \rightarrow \Theta} \right] + \left. \frac{\partial \tilde \omega_{Tj}}{\partial \Phi} \right|_{\Phi \rightarrow \Theta} \chi \cos (\omega_\Phi t) + \frac{\chi^2}{4} \left. \frac{\partial^2 \tilde \omega_{Tj}}{\partial \Phi^2} \right|_{\Phi \rightarrow \Theta} \cos (2\omega_\Phi t).
\end{aligned}
\end{equation}
There is a similar expansion for the coupling $J^x$. In a frame rotating at the qubit frequencies for $\chi=0$, oscillating $\sigma^z$ terms and DC exchange coupling terms time average to zero. This means that the time averaged qubit frequencies and couplings becomes
\begin{subequations}
\begin{align}
	\bar\omega_{Tj}(\Phi(t)) 
	=& \; \tilde \omega_{Tj} (\Theta) - \frac{\chi^2}{4} \left. \frac{\partial^2 \tilde \omega_{Tj}}{\partial \Phi^2} \right|_{\Phi \rightarrow \Theta}, \\
	\bar J^x(\Phi(t), t)  =&  \left. \frac{\partial \tilde J^x}{\partial \Phi} \right|_{\Phi \rightarrow \Theta} \chi \cos (\omega_\Phi t) + \frac{\chi^2}{4} \left. \frac{\partial^2 \tilde J^x}{\partial \Phi^2} \right|_{\Phi \rightarrow \Theta} \cos (2\omega_\Phi t). \label{eq:barJx}
\end{align}
\end{subequations}
Because there is a drive-induced qubit shift, all $N$ qubits will acquire a
phase during the flux modulation pulse. This phase may be compensated after by applying single-qubit Z-gates. In a frame rotating with $\hat H_0$ (including the drive-induced shift) the Hamiltonian is
\begin{equation}
	\hat H = -\frac{\bar\Delta_{T1}}{2}\sigma_{T1}^z + \sum_{i=1}^n \frac{J_{i}^z}{2}\sigma^z_{T1}\sigma_i^z + \bar J^x(\Phi(t), t)\left(\sigma^+_{T1}\sigma^-_{T2} + \sigma^-_{T1}\sigma^+_{T2}\right).
\end{equation}
Now in order to fix the oscillation of the exchange coupling and create the controlled \iswap gate we require the flux frequency to be resonant with the phase when the target qubits are in the $|1\rangle$ state, i.e.,  $\omega_\Phi = \bar \Delta_{T1} + \sum_{i=1}^{n}J^z_i$. In a frame rotating with the diagonal the Hamiltonian takes the form
\begin{equation}
\begin{aligned}
	\hat H =& \bar J^x(\Phi(t), t) e^{i(\bar\Delta_{T1} - \sum_{ i=1}^n J^z_i\sigma_i^z) t}\left(\sigma^+_{T1}\sigma^-_{T2} + \sigma^-_{T1}\sigma^+_{T2}\right) \\
	=& \chi\left. \frac{\partial \tilde J^x}{\partial \Phi} \right|_{\Phi \rightarrow \Theta} |\tilde 1 \rangle \langle \tilde 1 |_C \otimes \left(\sigma^+_{T1}\sigma^-_{T2} + \sigma^-_{T1}\sigma^+_{T2}\right),
\end{aligned}
\end{equation}
where we have used to rotating wave approximation to remove all fast rotating terms, i.e., all terms other than the term where all control qubits are in the state $|1 \rangle$. Note that there is also a resonant coupling at $2\omega_\Phi = \bar \Delta_{T1} + \sum_{i=1}^{n}J^z_i$, in which case the exchange coupling is via the second order terms in \cref{eq:barJx}

\section{Realistic parameters}

Here we presents parameters for the circuit model in \cref{fig:fredkin}(a), which yields the desired gate model of \cref{fig:fredkin}(b). The parameters are found by calculating the gate model parameters presented in \cref{app:analCircuit} and then minimizing a cost function which returns a low value when the requirements of the gate model are met. The minimization is done with using the simplex method, with randomized starting points, since many solutions exists. In order to judge the quality of the circuit parameters we also calculate the relative anharmonicity of the two-level systems, i.e., the difference between the 01 and the 12 transition, and the ratio between the effective Josephson energy and the effective capacitive energy.

The circuit parameters obtained are presented in \cref{tab:circuitParams} and corresponding parameters can be seen in \cref{tab:gateParams}. In \cref{tab:qualiParams} we present quality parameters (anharmonicities and effective Josephson junction and effective capacitance ratios) for the corresponding models. The parentheses in \cref{tab:gateParams,tab:qualiParams} indicates the error on the parameters, when assuming fabrication error on the circuit parameters in \cref{tab:circuitParams}. The errors are found using Monte Carlo simulations, where circuit parameters are drawn from a normal distribution centered around the experimental values presented in \cref{tab:circuitParams}, with one standard deviation corresponding to a 5\% error in the gate parameter. This correspond to 95\% of the drawn samples being within a 10\% error. The resulting errors in \cref{tab:gateParams,tab:qualiParams} corresponds one standard deviation.

Note that while it might look problematic that exchange coupling, $J^x$, is not significantly lower than the longitudinal coupling, $J^z$, this is not the case as the coupling of the gate is due to the first or second order term in \cref{eq:barJx}. This means that we can lower the exchange coupling to a desirable level when operating the gate. It is, however, important that $J^x$ is significantly lower than the detuning of the target qubit, which is indeed the case in all cases.

Note that all qubits have an anharmonicity above 2 \%, which is sufficient in order to suppress higher order levels in the anharmonic oscillator \cite{Koch2007}. We also note that the ratio between effective Josephson energy and effective capacitance are above 70, which is sufficient in order to suppress significant charge noise \cite{Koch2007}.

\begin{table}
	\caption{Circuit parameters for implementing possible controlled \iswap gates. Since the circuit parameter space is rather large we have several possible solutions; some, but far from all, possible solutions are show in the table. Here $E_1$, $E_{T1}$, $E_{T2}$, $E_{TB}$, and $E_{z,i}$ indicate the Josephson junction of the control qubit, target qubits, the tunable bus qubit, and the coupling between the target qubit and the control qubit, respectively; $C_1$, $C_{T1}$, $C_{T2}$, $C_{TB}$, $C_{z}$, and $C_{x}$ indicate the capacitance of the control qubit, the target qubits, and the couplings between them, respectively. Corresponding gate parameters can be seen in \cref{tab:gateParams}.}
	\label{tab:circuitParams}
	\begin{tabular}{c|ccccccccccc}
		\toprule
		& $E_1$ & $E_{T1}$ & $E_{T2}$ & $E_{TB}$ & $E_{z,1}$ & $C_1$ & $C_{T1}$ & $C_{T2}$ & $C_{TB}$ & $C_{z}$ & $C_{x}$ \\
		\# & $[2\pi \si{\GHz}]$ & $[2\pi \si{\GHz}]$ & $[2\pi \si{\GHz}]$ & $[2\pi \si{\GHz}]$ & $[2\pi \si{\GHz}]$ & $[\si{\femto\farad}]$ & $[\si{\femto\farad}]$ & $[\si{\femto\farad}]$ & $[\si{\femto\farad}]$ & $[\si{\femto\farad}]$ & $[\si{\femto\farad}]$ \\
		\toprule
		1 & $44.22$ & $12.63$ & $11.44$ & $0.41$ & $14.70$ & $1.00$ & $1.00$ & $68.32$ & $100.00$ & $54.26$ & $27.16$ \\ 
		2 & $24.88$ & $21.28$ & $53.12$ & $1.46$ & $9.18$ & $41.84$ & $31.47$ & $12.39$ & $27.92$ & $6.74$ & $7.46$ \\ 
		3 & $8.10$ & $33.93$ & $44.86$ & $5.09$ & $58.93$ & $16.63$ & $16.93$ & $29.69$ & $8.35$ & $10.01$ & $1.00$ \\ 
		4 & $0.09$ & $0.01$ & $13.27$ & $0.41$ & $33.39$ & $8.28$ & $19.38$ & $36.98$ & $99.30$ & $60.64$ & $35.75$ \\ 
		5 & $32.32$ & $20.14$ & $35.10$ & $0.51$ & $24.50$ & $4.80$ & $1.00$ & $1.00$ & $82.82$ & $79.21$ & $20.58$ \\ 
		6 & $20.48$ & $0.01$ & $17.14$ & $1.03$ & $50.65$ & $1.01$ & $23.85$ & $61.90$ & $39.50$ & $45.53$ & $10.27$ \\ 
		7 & $52.99$ & $45.93$ & $29.78$ & $1.04$ & $6.44$ & $8.21$ & $3.06$ & $27.90$ & $39.54$ & $70.97$ & $9.79$ \\ 
		8 & $29.58$ & $7.72$ & $21.40$ & $0.71$ & $28.64$ & $26.54$ & $32.89$ & $33.37$ & $56.54$ & $1.00$ & $16.55$ \\ 
		9 & $9.50$ & $26.92$ & $31.10$ & $2.90$ & $8.08$ & $72.77$ & $37.09$ & $30.97$ & $12.31$ & $49.03$ & $7.34$ \\ 
		10 & $10.77$ & $10.58$ & $17.51$ & $1.59$ & $16.70$ & $56.66$ & $7.18$ & $61.71$ & $23.92$ & $52.22$ & $10.16$ \\ 
		\toprule
	\end{tabular}
\end{table}

	\begin{table}
	\caption{Gate model parameters for implementing possible controlled \iswap gates corresponding to the circuit parameters in \cref{tab:circuitParams}. Since the circuit parameter space is rather large we have several possible solutions; some, but far from all, possible solutions are show in the table. Column 1-4 shows the dressed qubit frequencies. Column 5 and 6 shows the couplings seen in \cref{eq:Jz} and \cref{eq:Jx}. The parentheses indicates the error in the parameters when assuming an fabrication error of 10 \% on the circuit parameters in \cref{tab:circuitParams}.}
	\label{tab:gateParams}
	\begin{tabular}{c|cccccc}
		\toprule
		 & $\omega_1$ & $\omega_{T1}$ & $\omega_{TB}$ & $\omega_{T2}$ & $J^z$ & $J^x$ \\
		\# & $[2\pi \si{\GHz}]$ & $[2\pi \si{\GHz}]$ & $[2\pi \si{\GHz}]$ & $[2\pi \si{\GHz}]$ & $[2\pi \si{\MHz}]$ & $[2\pi \si{\MHz}]$\\
		\toprule 
1 & $\SI{16.61 +- 0.43}{}$ & $\SI{7.21 +- 0.20}{}$ & $\SI{1.08 +- 0.04}{}$ & $\SI{3.98 +- 0.12}{}$ & $\SI{-90.7 +- 5.2}{}$ & $\SI{8.2 +- 0.8}{}$ \\ 
2 & $\SI{9.63 +- 0.26}{}$ & $\SI{8.90 +- 0.37}{}$ & $\SI{3.82 +- 0.13}{}$ & $\SI{17.65 +- 0.52}{}$ & $\SI{-55.6 +- 3.4}{}$ & $\SI{20.7 +- 2.2}{}$ \\ 
3 & $\SI{16.37 +- 0.50}{}$ & $\SI{18.56 +- 0.43}{}$ & $\SI{13.37 +- 0.46}{}$ & $\SI{14.85 +- 0.51}{}$ & $\SI{-299.6 +- 10.7}{}$ & $\SI{27.3 +- 19.3}{}$ \\ 
4 & $\SI{8.45 +- 0.31}{}$ & $\SI{4.82 +- 0.18}{}$ & $\SI{1.06 +- 0.04}{}$ & $\SI{4.51 +- 0.12}{}$ & $\SI{-148.9 +- 4.1}{}$ & $\SI{11.8 +- 1.0}{}$ \\ 
5 & $\SI{14.66 +- 0.32}{}$ & $\SI{10.13 +- 0.26}{}$ & $\SI{1.34 +- 0.05}{}$ & $\SI{11.54 +- 0.40}{}$ & $\SI{-123.9 +- 6.3}{}$ & $\SI{11.1 +- 0.9}{}$ \\ 
6 & $\SI{17.69 +- 0.47}{}$ & $\SI{10.28 +- 11.09}{}$ & $\SI{2.70 +- 0.09}{}$ & $\SI{5.82 +- 0.22}{}$ & $\SI{-252.8 +- 7.3}{}$ & $\SI{14.3 +- 1.6}{}$ \\ 
7 & $\SI{18.51 +- 0.46}{}$ & $\SI{15.92 +- 0.45}{}$ & $\SI{2.73 +- 0.09}{}$ & $\SI{10.18 +- 0.25}{}$ & $\SI{-41.1 +- 2.9}{}$ & $\SI{13.6 +- 1.3}{}$ \\ 
8 & $\SI{15.84 +- 0.55}{}$ & $\SI{7.09 +- 0.18}{}$ & $\SI{1.88 +- 0.06}{}$ & $\SI{7.30 +- 0.20}{}$ & $\SI{-141.6 +- 5.9}{}$ & $\SI{12.2 +- 1.3}{}$ \\ 
9 & $\SI{4.71 +- 0.14}{}$ & $\SI{2.90 +- 2.81}{}$ & $\SI{7.62 +- 0.22}{}$ & $\SI{10.64 +- 0.35}{}$ & $\SI{-35.6 +- 1.9}{}$ & $\SI{16.7 +- 110.7}{}$ \\ 
10 & $\SI{6.07 +- 0.18}{}$ & $\SI{7.41 +- 0.18}{}$ & $\SI{4.19 +- 0.12}{}$ & $\SI{5.87 +- 0.20}{}$ & $\SI{-93.5 +- 3.9}{}$ & $\SI{47.3 +- 7.5}{}$ \\ 
		\toprule
	\end{tabular}
\end{table}

	\begin{table}
	\caption{Quality parameters for implementing possible controlled \iswap gates corresponding to the circuit parameters in \cref{tab:circuitParams}. $\alpha$ are the relative anharmonicities of the qubits, while $E^J/E^C$ are the ratios between the effective Josephson energy and effective capacitive energy. The parentheses indicates the error in the parameters when assuming an fabrication error of 10 \% on the circuit parameters in \cref{tab:circuitParams}.}
	\label{tab:qualiParams}
	\begin{tabular}{c|cccccccc}
		\toprule
		\# & $\alpha_1$ [\%] & $\alpha_{T1}$ [\%] & $\alpha_{TB}$ [\%] & $\alpha_{T2}$ [\%] & $E^{J}_1/E^{C}_1$ & $E_{T1}^J/E_{T1}^C$ & $E_{TB}^J/E_{TB}^C$ & $E_{T2}^J/E_{T2}^C$\\
		\toprule 
1 & $\SI{-2.1 +- 0.1}{}$ & $\SI{-2.5 +- 0.1}{}$ & $\SI{-2.1 +- 0.1}{}$ & $\SI{-2.0 +- 0.1}{}$ & $\SI{85.1 +- 4.0}{}$ & $\SI{77.1 +- 4.5}{}$ & $\SI{73.6 +- 4.9}{}$ & $\SI{71.4 +- 3.9}{}$ \\ 
2 & $\SI{-2.1 +- 0.1}{}$ & $\SI{-2.1 +- 0.1}{}$ & $\SI{-2.1 +- 0.1}{}$ & $\SI{-2.0 +- 0.1}{}$ & $\SI{83.9 +- 4.5}{}$ & $\SI{81.4 +- 3.9}{}$ & $\SI{74.3 +- 4.9}{}$ & $\SI{73.7 +- 3.9}{}$ \\ 
3 & $\SI{-2.6 +- 0.1}{}$ & $\SI{-2.1 +- 0.1}{}$ & $\SI{-2.1 +- 0.1}{}$ & $\SI{-2.1 +- 0.1}{}$ & $\SI{80.2 +- 4.7}{}$ & $\SI{120.6 +- 6.2}{}$ & $\SI{74.3 +- 5.1}{}$ & $\SI{73.3 +- 4.6}{}$ \\ 
4 & $\SI{-2.6 +- 0.1}{}$ & $\SI{-2.1 +- 0.1}{}$ & $\SI{-2.0 +- 0.1}{}$ & $\SI{-2.0 +- 0.1}{}$ & $\SI{76.4 +- 4.3}{}$ & $\SI{164.4 +- 10.3}{}$ & $\SI{74.4 +- 4.9}{}$ & $\SI{72.4 +- 3.9}{}$ \\ 
5 & $\SI{-2.1 +- 0.1}{}$ & $\SI{-2.1 +- 0.1}{}$ & $\SI{-2.0 +- 0.1}{}$ & $\SI{-2.0 +- 0.1}{}$ & $\SI{93.3 +- 4.1}{}$ & $\SI{104.9 +- 5.7}{}$ & $\SI{74.7 +- 5.1}{}$ & $\SI{74.3 +- 4.5}{}$ \\ 
6 & $\SI{-2.3 +- 0.1}{}$ & $\SI{-2.1 +- 0.3}{}$ & $\SI{-2.0 +- 0.1}{}$ & $\SI{-2.0 +- 0.1}{}$ & $\SI{85.7 +- 3.9}{}$ & $\SI{117.1 +- 7.0}{}$ & $\SI{74.4 +- 5.0}{}$ & $\SI{72.4 +- 4.2}{}$ \\ 
7 & $\SI{-2.1 +- 0.1}{}$ & $\SI{-2.1 +- 0.1}{}$ & $\SI{-2.1 +- 0.1}{}$ & $\SI{-2.0 +- 0.1}{}$ & $\SI{76.9 +- 3.7}{}$ & $\SI{79.6 +- 3.8}{}$ & $\SI{73.9 +- 5.0}{}$ & $\SI{72.1 +- 3.9}{}$ \\ 
8 & $\SI{-2.2 +- 0.1}{}$ & $\SI{-2.1 +- 0.1}{}$ & $\SI{-2.1 +- 0.1}{}$ & $\SI{-2.0 +- 0.1}{}$ & $\SI{82.7 +- 5.0}{}$ & $\SI{123.6 +- 6.7}{}$ & $\SI{73.9 +- 4.8}{}$ & $\SI{72.2 +- 3.9}{}$ \\ 
9 & $\SI{-2.1 +- 0.1}{}$ & $\SI{-4.2 +- 0.7}{}$ & $\SI{-2.0 +- 0.1}{}$ & $\SI{-2.0 +- 0.1}{}$ & $\SI{88.7 +- 4.8}{}$ & $\SI{144.9 +- 7.1}{}$ & $\SI{74.5 +- 4.6}{}$ & $\SI{71.9 +- 4.0}{}$ \\ 
10 & $\SI{-2.1 +- 0.1}{}$ & $\SI{-2.4 +- 0.1}{}$ & $\SI{-2.1 +- 0.1}{}$ & $\SI{-2.0 +- 0.1}{}$ & $\SI{105.4 +- 5.7}{}$ & $\SI{75.7 +- 3.4}{}$ & $\SI{74.3 +- 4.7}{}$ & $\SI{73.3 +- 4.2}{}$ \\ 
		\toprule
	\end{tabular}
\end{table}

\end{document}